\documentclass[journal]{IEEEtran}
\usepackage{cite}
\usepackage{verbatim}
\usepackage{epstopdf}
\usepackage{graphicx}
\usepackage{textcomp}
\usepackage[margin={0.7in}]{geometry}
\usepackage{amsmath,amssymb,amsfonts,bm}
\usepackage{booktabs}
\usepackage{epsf}
\usepackage{tikz}
\usepackage{psfrag}
\usepackage{pstool}
\usepackage{stfloats}

\usepackage{url}

\usepackage[dvips]{epsfig}
\usepackage{amsthm}
\usepackage{color}
\usepackage[usenames,dvipsnames]{pstricks}
\usepackage{lettrine}
\usepackage[dvips]{epsfig}
\usepackage{pst-grad} % For gradients
\usepackage{pst-plot} % For axes
\usepackage{epsfig}
\usepackage{enumerate}
\usepackage{amsbsy}
\usepackage{amssymb}
\usepackage{amsthm}
\usepackage{amscd}
\usepackage{float}
\usepackage[caption = false]{subfig}
\usepackage{color}
\usepackage[numbers,sort&compress,square]{natbib}

\usepackage{listings}
\usepackage{stackrel}
\usepackage{authblk}
\usepackage{dsfont}
\usepackage{algorithm}
\usepackage[noend]{algpseudocode}
\makeatletter
\newcommand{\removelatexerror}{\let\@latex@error\@gobble}
\makeatother

\usepackage{array}
\newcolumntype{L}[1]{>{\raggedright\let\newline\\\arraybackslash\hspace{0pt}}m{#1}}
\newcolumntype{C}[1]{>{\centering\let\newline\\\arraybackslash\hspace{0pt}}m{#1}}
\newcolumntype{R}[1]{>{\raggedleft\let\newline\\\arraybackslash\hspace{0pt}}m{#1}}

\ifCLASSINFOpdf

\fi

\def\BibTeX{{\rm B\kern-.05em{\sc i\kern-.025em b}\kern-.08em
    T\kern-.1667em\lower.7ex\hbox{E}\kern-.125emX}}
\begin{document}
\title{Self-Evolving Integrated Vertical Heterogeneous Networks}


\author{Amin Farajzadeh, \and Mohammad G. Khoshkholgh, \and Halim Yanikomeroglu, and \and Ozgur Ercetin
\thanks{This work was supported by Huawei Canada Co. Ltd.\\
\indent A. Farajzadeh, M. G. Khoshkholgh, and H. Yanikomeroglu are with the Department of Systems
and Computer Engineering, Carleton University, Ottawa, ON K1S 5B6,
Canada (e-mail: aminfarajzadeh@sce.carleton.ca; m.g.khoshkholgh@gmail.com; halim@sce.carleton.ca)\\
\indent O. Ercetin is with the Faculty of Engineering and Natural Sciences, Sabanci
University, 34956 Istanbul, Turkey (e-mail: oercetin@sabanciuniv.edu).
}}

\maketitle

%\begin{abstract}

%\end{abstract}


% \fi
\IEEEpeerreviewmaketitle
\begin{abstract}
6G and beyond networks tend towards fully intelligent and adaptive design in order to provide better operational agility in maintaining universal wireless access and supporting a wide range of services and use cases while dealing with network complexity efficiently. Such enhanced network agility will require developing a \textit{self-evolving} capability in designing both the network architecture and resource management to intelligently utilize resources, reduce operational costs, and achieve
the coveted quality of service (QoS). To enable this capability, the necessity of considering an integrated vertical heterogeneous network (VHetNet) architecture appears to be inevitable due to its high inherent agility. Moreover, employing an intelligent framework is another crucial requirement for self-evolving networks to deal with real-time network optimization problems. Hence, in this work, to provide a better insight into network architecture design in support of self-evolving networks, we highlight the merits of integrated VHetNet architecture while proposing an intelligent framework for self-evolving integrated vertical heterogeneous networks (SEI-VHetNets). The impact of the challenges associated with SEI-VHetNet architecture, on network management is also studied considering a generalized network model. Furthermore, the current literature on network management of integrated VHetNets along with the recent advancements in artificial intelligence (AI)/machine learning (ML) solutions are discussed. Accordingly, the core challenges of integrating AI/ML in SEI-VHetNets are identified. Finally, the potential future research directions for advancing the autonomous and self-evolving capabilities of SEI-VHetNets are discussed.

\end{abstract}
%\begin{table*}[!t]
%\caption{List of Acronyms.}
%\begin{center}
%\begin{tabular}{c|c||c|c}
%\toprule %
%\textbf{Acronyms}&  %\textbf{Definitions}&\textbf{Acronyms}&  %\textbf{Definitions}
%\\[.3\normalbaselineskip]
%\hline
%\end{tabular}
%\end{center}
%\end{table*}
\section{Introduction}
\subsection{Motivation}
Contemporary networks are an amalgamation of distinct terrestrial, aerial, and space/satellite platforms or tiers. These tiers operate independently and under loose coordination, with human supervision as a standard and essential part of their operations~\cite{intro-n1}. Continuous monitoring and intervention by expert engineers limit the capability to meet new and unforeseen requirements, maintain universal wireless access, and support novel use-cases and services yet to be conceptualized in 6G networks~\cite{6g-1}. To fulfill these, two necessary approaches are needed: 1) revisit the current network architecture and aim to design an integrated network that consists of all terrestrial, aerial, and space/satellite tiers; 2) enable fully autonomous coordination and management in different aspects of the network, including network architecture and resource management. Recently, integrated vertical heterogeneous network (VHetNet) architecture has been introduced to combine all the vertical tiers~\cite{VHETNET}. Here vertical refers to a physical characteristic of such networks, e.g., altitude heterogeneity of the network layers or tiers. Hence, in this work, the fully intelligent and adaptive coordination and management of integrated VHetNets is advocated. Unlike terrestrial networks for which there has been steady progress in the network management policies~\cite{self-healing}, integrated scenarios of aerial, terrestrial, and satellite networks, i.e., integrated  VHetNets, have yet to be thoroughly studied. Indeed, integrated VHetNets face many challenges despite all their advantages when it comes to full integration and coordination between the vertical tiers, and network resource management~\cite{II-9}. In a fully coordinated integrated VHetNet, available resources need to be assigned efficiently across vertical tiers based on the user demands/needs. Also, the network topology should be managed continuously in an adaptive fashion based on the use-case, requested services, and the future predicted demands~\cite{Wael}. Such coordinated and adaptive network management is unprecedented and cannot be accomplished by existing solutions and technologies such as network virtualization~\cite{intro-n2}. Hence, to tackle these challenges in integrated VHetNets, it is necessary to develop an intelligent framework that enables fully autonomous and adaptive network management across all vertical tiers, while ensuring full coordination and integration~\cite{Intro2}.

In recent literature, self-organizing networks (SONs) and machine learning (ML) enabled SONs have been introduced to enable intelligent network management~\cite{son01}-\cite{son02}. In these works, the focus is on rule-based configuration, optimization, and adaptation of an existing terrestrial network~\cite{Intro3}-\cite{intro-n4}. However, to bring intelligence to integrated VHetNets, it will require a paradigm shift from conventional SON architecture to adapt the functionalities of integrated VHetNets and their topology to specific environments in self-evolving ways, allowing these networks to respond to a wide range of user demands and perform under dynamic and complex conditions. Such self-evolving capability will transform network management from self-organizing into a continuous, fully automatic, and intelligently evolving entity~\cite{survy20}. A self-evolving structure is expected to drive network management
from self-organization to continuous and automatic
evolvement, where even management policies can
be self-adjusted, to automatically react to unknown
environments and triggers, requiring self-adaptive
and resilient learning mechanisms. Specifically, continuous and automatic configuration and coordination targets real-time automated initial network parameter configuration and auto-connectivity. Moreover, self-evolving networks aim to self-manage a network of networks that spans across multiple operators, tiers, and ecosystems (e.g., satellite, aerial, and terrestrial
networks). It is also expected to support continuous, adaptive, and automated self-optimization aiming to optimize certain aspects and operations such as coverage and capacity, interference, handover settings, and energy savings. Hence, unlike SONs, self-evolving networks are believed to be capable of making autonomous management decisions across multiple vertical tiers, making them suitable for integrated VHetNets. In addition, the self-evolving capability will improve overall network performance across each vertical tier and will manage coordination among the several entities in the future integrated networks~\cite{survy20}. This capability will also consider the provision, optimization, and management of all aspects of network management, including communication, computational, control, hardware, and software resources. TABLE I provides a comparison
between the characteristics of SON and self-evolving networks. %Ideally, any outstanding interactions and interconnections among these entities are commonly designed and investigated in isolation. 

The main tool to enable self-evolving capability is artificial intelligence (AI) which has been considered the science of training machines to perform human tasks~\cite{AI-n1}. 
%There are many applications for AI, including robotic vehicles, speech recognition, machine translation, and, recently, wireless communications~\cite{AI-n2}.
A specific subset of AI involves training machines on how to learn, which constitutes a new framework known as Machine Learning (ML)~\cite{intro-ML1}. In this context, ML can provide solutions in scenarios where a massive number of devices simultaneously require access to the network's resources in a dynamic, heterogeneous, and unpredictable way (e.g., in IoT communications)~\cite{intro-ML2}. ML is expected to be a key solution for empowering self-evolving networks in order to configure, coordinate, manage, and optimize beyond 5G network functions in an automated, adaptive, and continuous fashion so as to achieve full coordination and integration, and full radio access network automation across multiple vertical tiers.
%In general, AI/ML techniques can be classified into several types: supervised learning, semi-supervised learning, unsupervised learning, and reinforcement learning (RL)~\cite{AI-n3}. 
In recent studies~\cite{intro-AI11}-\cite{intro-AI13}, ML has mainly been considered for resolving complications in the design of communications protocols which subsequently degrade the communication efficiency and network overall performance. The use of machine learning methods to develop communication systems has demonstrated its effectiveness~\cite{into-AI1},~\cite{intro-AI2}. Recent applications of various ML techniques
%, such as reinforcement learning (RL), 
have been studied in coexisting terrestrial-aerial wireless networks~\cite{Intro5}. In most of these works~\cite {Intro7}-\cite{Intro10}, the main objective has been to find the best strategy for network resource management on the basis of local observations using modern ML methods~\cite{Intro11}.
%Specifically, they aim to improve the performance of one (or joint) network management aspect such resource allocation and UAV trajectory design~\cite{Intro12}.
Although the idea of employing AI/ML techniques to solve network optimization problems is gaining momentum, few studies exist in the literature on how AI/ML can ensure the coordination, connectivity, and integration of networks~\cite{AI-n5}. As network architecture becomes increasingly heterogeneous, the deployment and integration of ML are becoming increasingly necessary due to the higher complexity of infrastructure and higher diversity of associated devices and resources~\cite{AI-n6}. In integrated VHetNets, which are highly heterogeneous and composed of several vertical tiers, the integration of AI/ML is a crucial requirement. In these networks, AI/ML can be employed for designing an intelligent mechanism to develop algorithms for network optimization while also guaranteeing full coordination, connectivity, and integration among tiers~\cite{AI-n7}. Furthermore, self-evolving capabilities require an intelligent framework that relies heavily on AI/ML, since these capabilities require real-time, online, and joint decisions or operations in different aspects of network management across vertical tiers~\cite{survy20}. Hence, the intelligence enabled by ML algorithms can work as the core of the self-evolving structure in an integrated VHetNet architecture and it will be powered by both the communications environment's collected data (e.g., spatial and temporal traffic distributions, user demands, and mobility patterns) and external source improvements such as novel technologies, emerging network components, and advanced
communication services.

In considering an integrated VHetNet with self-evolving capability which we refer to as a self-evolving integrated (SEI)-VHetNet, the following objectives should be addressed: 
\begin{itemize}
    \item What are the unique attributes, key components, and challenges of a typical SEI-VHetNet architecture? What are the main differences between SON and self-evolving networks? (Section II)
    \item What are the impacts of challenges associated with integrated VHetNet architecture, on network management policies and, accordingly, the formulation of network management optimization problems? (Section III)
    \item How to design an intelligent framework to enable self-evolving capability in integrated VHetNets, and analyze and deal with associated real-time network management problems? (Section IV)
  \item Why do intelligent solution methods are of necessary for coordination between the vertical tiers and satisfy the self-evolving requirements? (Section IV)
  \item What are the most effective approaches in terms of performance metrics, such as delay, accuracy, precision time, network throughput, etc., for addressing the challenges of ML integration in SEI-VHetNets? (Section V)  
\end{itemize}
\subsection{Existing Related Surveys}
A number of surveys exist in the literature on various aspects of wireless networks with aerial nodes. TABLE I provides a classification along with the summaries of these works. Although these surveys address important communication problems related to various aspects of so-called unmanned aerial vehicle (UAV)-enabled communications, they do not investigate the management aspect of these networks. Also, only a few considered AI/ML as an enabler for vertical networks.  Finally, only one work addressed an integrated tiered architecture.  Based on these studies, major challenges appear as the effect of high mobility, dynamic vertical topology, full integration complexity, stability and reliability of platforms, and safety.
%%%%%%%%
%
\begin{figure*}[!t]
    \centering
    \includegraphics[width=175mm,scale=10]{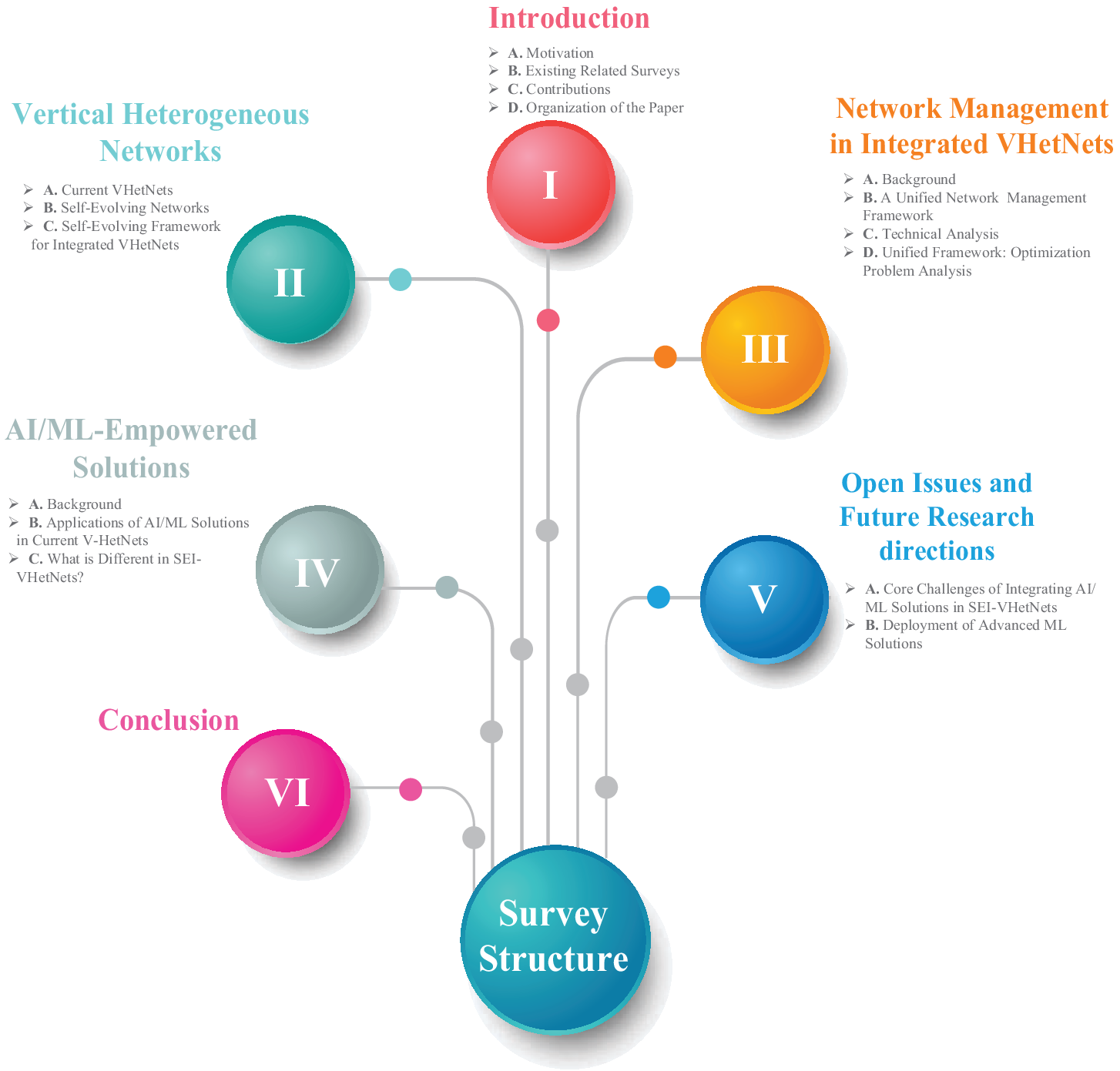} 
    \caption{The structural diagram of this paper.}
    \label{fig1}
\end{figure*}
%%%%%%%%%%%%%%%%%%%%%%%%%%%
%%%%%%%%%%%%%%%%%%
\begin{table*}[!t]\label{tab1}
\caption{Summary and classification of existing related works.}
\begin{center}
\begin{tabular}{L{0.35cm}|L{1.3cm}|L{1.3cm}|L{1cm}|L{1.1cm}|L{10.7cm}}
\toprule %
\textbf{Ref.}&  \textbf{UAV-enabled wireless networks}& \textbf{Integrated three-tier architecture}& \textbf{AI/ML discussion}&\textbf{Network management analysis}&\textbf{Main Contribution}%&\textbf{Main contents}
\\[.3\normalbaselineskip]
\hline
\cite{survy20}&$\times$&\checkmark&\checkmark&$\times$& A visionary and descriptive paper on self-evolving networks powered by AI/ML for making intelligent, adaptive, and automated decisions.
\\[.3\normalbaselineskip]
\hline
\cite{surv1} &\checkmark&$\times$&$\times$&$\times$&Practical issues and solutions for the integration of UAVs as base stations in cellular networks.\\[.3\normalbaselineskip]
\hline
\cite{surv2} &\checkmark&$\times$&$\times$&$\times$& Overview of power
control in UAV-supported UDNs. Four representative scenarios are investigated, including aerial base stations, mobile relays, energy transfers, and caching. \\[.3\normalbaselineskip]
\hline
\cite{surv3} &\checkmark&$\times$&$\times$&$\times$&Examination of mobility models and routing techniques in flying ad hoc networks (FANETs).\\[.3\normalbaselineskip]
\hline
\cite{survy4a}&\checkmark&$\times$&$\times$&$\times$& Classification of communication, and application architectures, and existing routing protocols for flying ad hoc networks (FANETs).\\[.3\normalbaselineskip]
\hline
\cite{surv5}  &\checkmark&$\times$&$\times$&$\times$& Review of recent research on UAV communications integration with various 5G technologies at the physical and network layers, including communication, computing, and caching.
\\[.3\normalbaselineskip]
\hline
\cite{surv6}&\checkmark&$\times$&$\times$&$\times$& Discussion of UAV-based 3D cellular network architecture including network planning, latency-minimal 3D cell association, 3D UAV placement, and frequency planning for UAV base stations. \\[.3\normalbaselineskip]
\hline
\cite{surv6(1B)} &\checkmark&$\times$&$\times$&$\times$&Survey of routing, seamless handover, and energy efficiency aspects of UAV communications. \\[.3\normalbaselineskip]
\hline
\cite{surv8} &\checkmark&$\times$&$\times$&$\times$&Overview of the characteristics and requirements of UAV networks for envisioned civil applications from a communications
and networking viewpoint. No technical discussion or discussion of AI. \\[.3\normalbaselineskip]
\hline
\cite{surv9}&\checkmark&$\times$&$\times$&$\times$&A study focusing specifically on two main use cases: aerial base stations and cellular-connected users.\\[.3\normalbaselineskip]
%\hline
%\cite{surv10} &&\\[.3\normalbaselineskip]
\hline
\cite{surv11} &\checkmark&$\times$&$\times$&$\times$& A survey and analysis of channel measurements and
models for both civil aeronautical communications and UAV
communications. \\[.3\normalbaselineskip]
\hline
\cite{surv12}&\checkmark&$\times$&$\times$&$\times$&A  survey of SDN and NFV-enabled UAV networks.\\[.3\normalbaselineskip]
 \hline
 \cite{surv13} &\checkmark&$\times$&$\times$&$\times$& An overview of available research on space-air-ground integrated networks. Also discusses cross-layer design, resource management, and allocation to system integration, network performance analysis, and optimization.\\[.3\normalbaselineskip]
 \hline\cite{surv14} &\checkmark&$\times$&$\times$&$\times$&A survey of current achievements, technical advantages, and challenges as well as potential applications, in the integration of 5G mmWave communications into UAV-assisted wireless networks.\\[.3\normalbaselineskip]
 \hline\cite{surv15} &\checkmark&$\times$&$\times$&$\times$&A tutorial overview of recent advances in UAV communications,
with an emphasis on integrating UAVs into new 5G and future
cellular networks. Also discusses performance analysis, evaluation, and optimization of UAV communication.\\[.3\normalbaselineskip]
\hline
\cite{surv16}&\checkmark&$\times$&$\times$&$\times$&
A survey of recent developments in game theory for UAV-aided wireless communications.\\[.3\normalbaselineskip]
  \hline
 \cite{survy17} &\checkmark&$\times$&$\times$&$\times$&A survey of UAV networks from a cyber-physical systems perspective. Also discusses the requirements, challenges, solutions, and advances of three cyber components, including communication, computation, and control. \\[.3\normalbaselineskip]
 \hline
 \cite{survy18}&\checkmark&$\times$&$\times$&$\times$&The current and future design challenges of multi-UAV systems in cyber-physical applications have been classified then elaborations on
each were given.\\[.3\normalbaselineskip]
\hline
\cite{survy19}&\checkmark&$\times$&$\times$&$\times$&A comprehensive survey of
air-to-ground propagation channels for UAVs.
\\[.3\normalbaselineskip]
\hline
\cite{survy21}&\checkmark&$\times$&\checkmark&$\times$&A visionary and descriptive survey of the potential of high altitude platform station (HAPS) systems for 6G networks.
\\[.3\normalbaselineskip]
\hline
\cite{SURVEY22}&$\times$&\checkmark&$\times$&\checkmark& An in-depth literature review focusing on the latest satellite communication contributions in terms of system aspects, air interface, medium access control techniques, and networking.
\\[.3\normalbaselineskip]
\hline
\cite{SURVEY23}&\checkmark&$\times$&\checkmark&$\times$& A comprehensive overview of ML techniques, including unsupervised, supervised, and federated learning
techniques that have been applied in UAV networks.
\\[.3\normalbaselineskip]
\hline
\cite{survey24}&\checkmark&\checkmark&$\times$&$\times$& A comprehensive overview of networked tethered flying platforms and their integration in non-terrestrial networks alongside free-flying platforms and satellites in 6G cellular technology.
\\[.3\normalbaselineskip]
\hline
\cite{survey25}&\checkmark&\checkmark&\checkmark&$\times$& An extensive review study on the evolution of non-terrestrial networks from 5G to 6G networks, their integration into terrestrial networks, and a new range of services and use cases.
\\[.3\normalbaselineskip]
\hline
\cite{survey26}&\checkmark&$\times$&\checkmark&$\times$& A comprehensive summary of ML techniques, including their applications and contributions towards more effective UAV network implementations.
\\[.3\normalbaselineskip]
\hline
\cite{R1-surv}&\checkmark&\checkmark&$\times$&\checkmark&A comprehensive comparison and review of different types of UAVs for communication services. It also discusses integrated air-to-ground heterogeneous network architecture and highlights its main characteristics and potential advantages.
\\[.3\normalbaselineskip]
\hline
This work&\checkmark&\checkmark&\checkmark&\checkmark&A visionary and technical study on the integrated VHetNet architecture enhanced by self-evolving capability, with the main emphasis on identifying the corresponding architecture-dependent challenges, network management requirements, and the employment of AI/ML techniques as key a component for self-evolving purposes.
\\[.3\normalbaselineskip]
\bottomrule 
\end{tabular}
\label{tab2}
\end{center}
\end{table*}
%%%%%%%%%%%%%%%%%%%%%%%%%%%%%%%%
%%%%%%%%%%%%%%%%%%%%%%%%%%
\begin{figure*}[!t]
    \centering
    \includegraphics[width=181mm,scale=10]{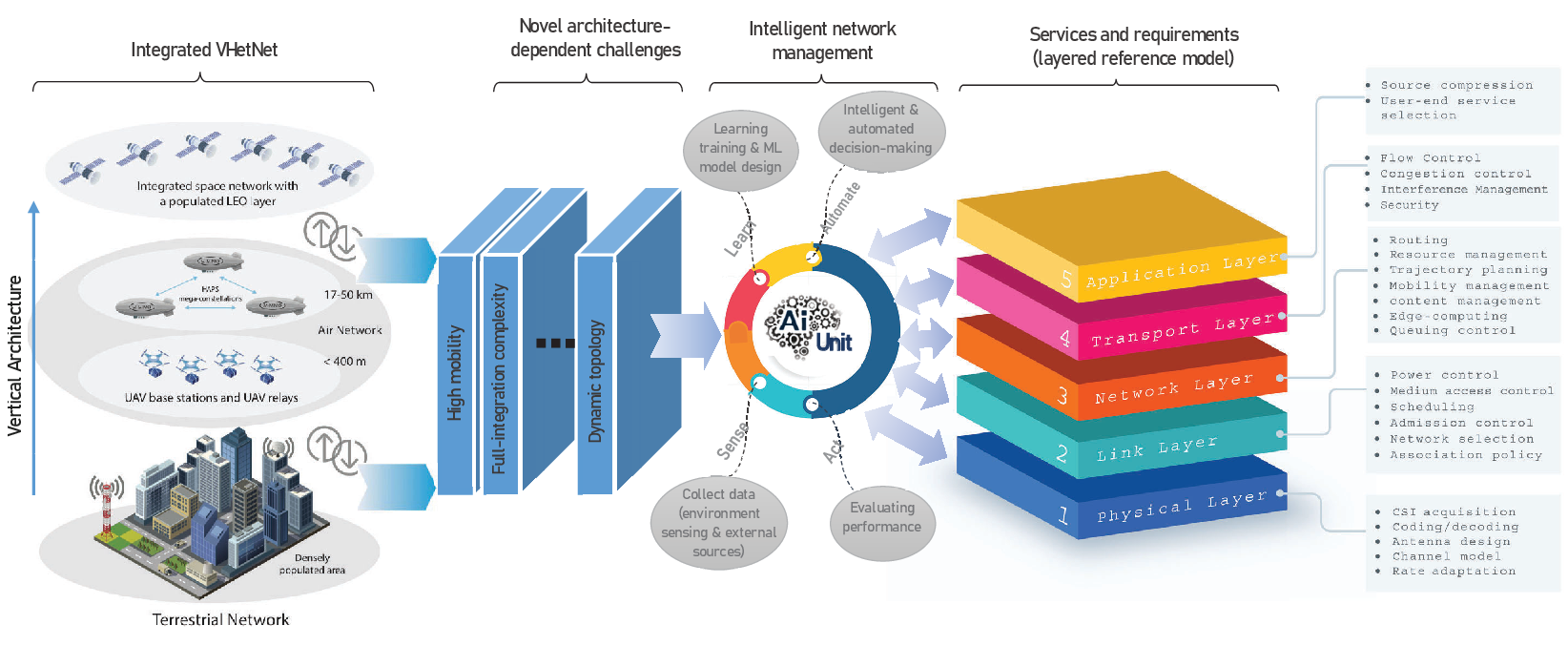}
 \caption{SEI-VHetNet: The big picture.}
    \label{fig2}
\end{figure*}
%%%%%%%%%%%%%%%%%
\subsection{Contributions}
In this work, we discuss a unified network management framework and examine its design issues and possible approaches suitable for current and future integrated vertical networks. We investigate vertical integration and coordination between three tiers with AI/ML-based network management.
%to ensure the full end-to-end intelligent coordination and integration across the tiers. 
The main contributions of this paper are as follows:
\begin{itemize}
    \item We conceptualize an architecture of SEI-VHetNets and identify open challenges to realize this architecture. %\textcolor{red}{(Section II)}
    \item We propose an intelligent framework for enabling self-evolving capabilities in integrated VHetNets.
    \item We investigate the impact of the challenges associated with integrated VHetNet architecture on network management considering  a  generalized  system  model. To fulfill this, we consider a unified network management framework and study the synergistic interaction of vertical tiers in SEI-VHetNets. % Provide a generalized problem formulation for each scenario. For instance, the correlation and interaction between the high mobility challenge and joint communication and control aspects, introduces a problem with a set of constraints imposed by high mobility challenge (ex. maximum speed constraint), communication aspect (ex. outage constraint), and control (ex. power constraint). In Fig.~\ref{fig3}, the unified network management framework and the corresponding interactions are illustrated. %\textcolor{red}{(Section III)}   
    \item We examine and provide a classification of various solution methods suitable for integrated VHetNets available in the literature. %\textcolor{red}{(Section III)}\\
    %(Some of the scenarios has not been studied yet so that we can make a better contribution at this point, at least by formulating the problems and discussing the potential AL/ML methods (Section V) in order to solve them.) 
    \item We investigate the suitability, effectiveness, and efficiency of modern ML methods for SEI-VHetNets. %\textcolor{red}{(Section IV)}
    %\item Performavaluation of ML methods applied to each interaction (i.e., joint optimization problem) in I-VHetNets under the consideration of key parameters such as accuracy and convergence speed. %\textcolor{red}{(Section V)}
    \item We provide our vision for potential future research directions for advancing the autonomous and self-evolving capabilities of SEI-VHetNets. %\textcolor{red}{(Section VI)}
\end{itemize}
\subsection{Organization of the Paper}
As shown in Fig.~\ref{fig1}, the rest of the paper is organized as follows. In Section II, we provide an overview of current VHetNets and the characteristics of self-evolving networks. In Section III, we discuss network management in integrated VHetNets and propose a unified network management system while also identifying new architecture-dependent challenges. Also in this section, we analyze network management optimization problems considering a generalized system model. In Section IV, we discuss AI/ML-based solutions for dealing with network optimization problems and examine the core challenges of AI/ML integration in SEI-VHetNets. Potential applications and future directions to enhance the SEI-VHetNets are studied in Section V. Finally, Section VI concludes the paper.
\section{Vertical Heterogeneous Networks}
%%%%%%%%%%%%%%%%%%%%%%%%%
\begin{table*}[!t]\label{tab2}
\caption{A comparison of SON and SE networks characteristics}
\begin{center}
\begin{tabular}{L{2cm}|L{6cm}|L{6cm}}
\toprule %
\textbf{Characteristics}&  \textbf{SON}& \textbf{SE networks}
\\[.3\normalbaselineskip]
\hline 
KPIs & KPIs are optimized within a single network at each tier individually. &Various and different KPIs are jointly optimized across multiple (e.g., all) vertical tiers.   \\[.3\normalbaselineskip]
\hline
Coordination, integration, and conflict management & Lacks full integration, coordination, and conflict avoidance among autonomic or self-established management functions of a SON.& Provides conflict-free and full coordination of multiple autonomic functions across vertical tiers. Full integration of all three tiers dues the ensured coordination. Supports multiple SONs that operate simultaneously in the same or interacting tiers or networks and operators across multi-tiers.\\[.3\normalbaselineskip]
\hline
AI/ML-based schemes deployment & Enables intelligence in both network edge and core. Intelligent schemes are deployed in a centralized or semi-centralized fashion within a single tier for associated operators.& Intelligent schemes are fully distributed across all tiers and operators, such that they are also adapted to various network management aspects.  \\[.3\normalbaselineskip]
\hline
Security $\&$
privacy level & Security and privacy are managed within a single network
or operator domain. & Security and privacy are managed across vertical tiers collaboratively and in a distributed manner for all operators. \\[.3\normalbaselineskip]
\hline
Network management, control, and
optimization&Self-establishment of 3GPP network function,
resource allocation, load balancing, interference
coordination, and random access optimization within
a single network or tier or operator. & The new functionality of SE networks brings in automated dynamic network expansion and
temporary network provision by controlling the mobility of infrastructure nodes, and
integrated cross-network and cross-tier optimization and resource management. \\[.3\normalbaselineskip]
\bottomrule 
\end{tabular}
\label{tab2}
\end{center}
\end{table*}
%%%%%%%%%%%%%%%%%%%%%%%%%%%%%%%%%
\subsection{Current VHetNets}
Due to their unique attributes such as high flexibility and mobility, aerial networks have already gained significant attention both in academia and industry~\cite{II-1}-\cite{II-5}. The 3rd generation partnership project (3GPP) has already initiated its activities regarding the non-terrestrial network (NTN), as defined in Technical Report 38.821~\cite{II-6}. Recently, a fully integrated VHetNet with three tiers, including terrestrial, aerial, and space/satellite, has been focused in literature~\cite{survy21} as a promising solution to provide feasible ubiquitous connectivity while supporting the requirements of unknown and possibly revolutionary applications. 
With the rapid proliferation of mobile wireless devices and accompanying data traffic, traditional terrestrial networks face difficulties in supporting data rates and ever-increasing strict delay requirements. Even more importantly, the cost of deploying terrestrial infrastructure is prohibitive in remote areas. To tackle these limitations and challenges, integrating current terrestrial networks with aerial networks along with satellite/space networks has become a necessity. In \cite{II-7}, it is stated that traditional communications between terrestrial nodes and satellite nodes (as base stations or relay nodes) suffer from several factors: 1) Traditional satellite communications are constrained by high path-loss attenuation, 2) satellites are located at various orbital altitudes causing significant propagation delays~\cite{sat11}, \cite{sat12}. To address these factors, aerial networks can be a promising solution, and they can work along with existing satellite and terrestrial networks \cite{II-8}. As shown in Fig.~\ref{fig2}, the three-tier architecture is composed of a satellite/space tier, an aerial tier, and a terrestrial tier. Aerial networks consist of two sub-tiers, including low-altitude platforms and high-altitude platforms (HAPs).
%Low altitude platform is located at relatively lower altitudes below $400$ m and. UAVs and drones fly at this platform and can be used to enhance network coverage and capacity by deploying a swarm of flying platforms that implement novel radio resource management techniques~\cite{}.
HAPs refer to airships and balloons operating in the stratosphere, at altitudes of $17$ Km to $20$ Km, depending on where the wind currents and turbulence are the lowest~\cite{hap1}. Therefore, HAPs can act as aerial base stations or relay nodes to improve the communication links between satellite stations and ground nodes and hence improve the overall network throughput. As the demand for comprehensive broadband services, global coverage, and ubiquitous access has grown, non-terrestrial networks can provide strong support for well-established terrestrial backhaul networks. Low altitude platforms operate below $400$ m and facilitate various civilian, commercial and governmental missions, as well as IoT applications, ranging from military and security operations to entertainment and telecommunications. Low-altitude platforms are mainly responsible for network optimization~\cite{Amin}. Unmanned aerial vehicles (UAVs) or drones are the major representatives of low-altitude platforms. UAVs are generally employed for short periods of time (up to several hours), allowing for the rapid deployment of a multi-hop communication backbone in challenging situations, such as public safety, search and rescue missions, surveillance, emergency communications in post-disaster situations, and so forth~\cite{UAV10}, \cite{Amin2}. Hence, this terrestrial-aerial-satellite integrated network architecture can play an important role in civilian life, with opportunities for improving wireless connectivity in general and for public safety and first responders in particular.
\subsection{Self-Evolving Networks}
To envision integrated VHetNets capable of adapting dynamically to changing topologies and network management policies, we need to first identify the main characteristics of self-evolving networks and their differences from self-organizing networks SONs. We then discuss a potential network structure for SE networks. In spite of the 3GPP-introduced progressive concept of SONs in 4G and 5G documents, e.g., to automate 
network management, several challenges/shortcomings may remain as the SONs hardly harness/provide sufficient agility for dealing with the immense levels of complexity, heterogeneity, and mobility in the envisioned beyond-5G integrated networks~\cite{Intro3}-\cite{R1SON}. Integrated VHetNets with self-evolving capabilities are expected to provide end-to-end intelligent and closed-loop network automation that is not limited to optimizing network configuration parameters but can reach the level of automatically forming a temporary communications network (i.e., through mobile and agile base stations) to fulfill the demands of specific regions for certain durations. In the following, we identify the main characteristics of self-evolving networks, taking integrated VHetNet architecture into account. Moreover, In TABLE II, we provide a summary comparison between the main characteristics of SON and self-evolving concepts.
\begin{itemize}
    \item {\bf{Distributed and Intelligent Decision-Making:}} By facilitating integrated VHetNets with self-evolving capabilities, there will be a set of self-evolving (SE) units that collaborate to exchange specific or general information. An SE unit will be able to create, organize, control, manage, and sustain itself autonomously based on its own network and environmental observations along with the exchanged information from other SE units. This will create high adaptability to changes in the network environment and increase the scalability, robustness, and fault tolerance.
    \item  {\bf{Fully Dynamic, 3D, and Agile Topology:}} Considering the architecture of integrated VHetNets, due to the 3D topology of each node at any vertical tier, everything in the network can move, including users and base stations (e.g., UAV, HAP, or satellite). Integrating self-evolving frameworks in such highly dynamic topologies will allow the forming, splitting, and slicing of networks based on changes in user demands and environmental conditions (e.g., link quality). Agility and flexibility of a self-evolving framework can prevent over-engineering or excessive densification of terrestrial networks, such that better network throughput, data rate, or coverage can be achieved for a wide range of applications. SE units will be able to intelligently, adaptively, and automatically manage network resources and adjust the network topology to cope with variable user demands and random changes in network environment status. 
    \item {\bf{Seamless Connectivity:}} 
    SEI-VHetNets are expected to support ubiquitous connectivity and massive network traffic while providing various services across all vertical tiers including terrestrial, aerial, and space tiers. The integration of an SE framework will ensure full integration and coordination and subsequently, seamless connectivity between these tiers since network resource management along with topology adjustment will be controlled collaboratively by SE units.
 \item {\bf{Distributed Multi-Level Computing and Caching:}} Due to the demand for high performance of remote resource access in transparent computing, there is a requirement in SEI-VHetNets to design an adaptive multi-level cache framework to provide a global cache resource scheduling and to alleviate the network latency. Existing cache frameworks in CPU and web systems cannot be applied simply because the remote resource access architecture needs to be extended to support multi-level cache, and the ways that resources are accessed in transparent computing require specific designs. To tackle these challenges, the SE framework can manage the collaboration among smart devices, e.g., autonomous vehicles, and with the edge computing nodes at any vertical tier to achieve distributed intelligent learning and decision-making. The computational and caching resources not only support terrestrial/aerial/space node applications, but they also support the intelligent automation functionality in SE networks. 
\end{itemize}
\subsection{Self-Evolving Framework for Integrated VHetNets}
\begin{figure*}[!t]
    \centering
    \includegraphics[width=150mm,scale=10]{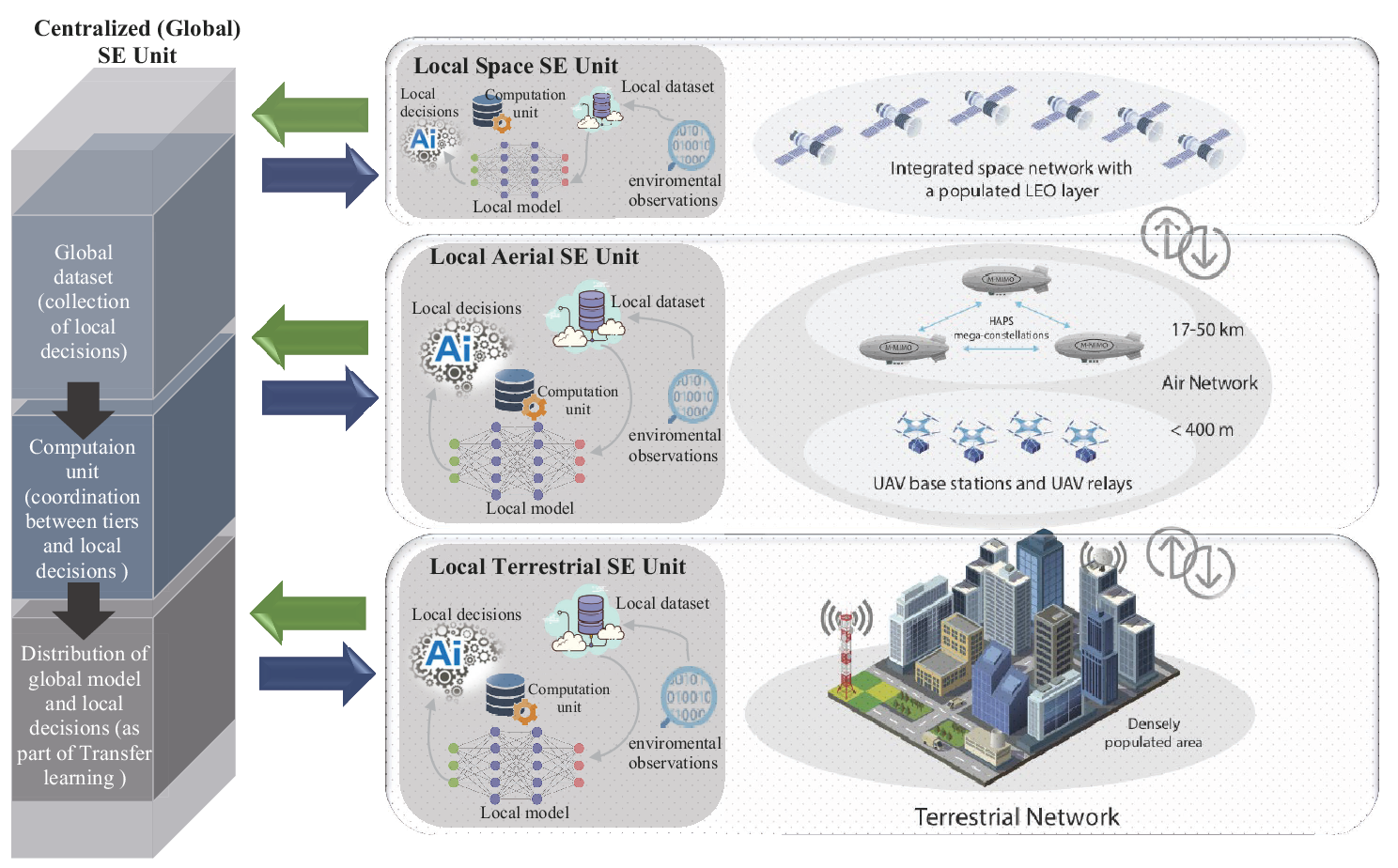} 
    \caption{Proposed intelligent framework for SEI-VHetNets. A key component to support this framework is a distributed AI approach (e.g., federated learning and transfer learning) to ensure full integration and coordination.}
    \label{fig5}
\end{figure*}

In Fig.~\ref{fig5}, we show the SE structure and its interaction with future integrated networks (e.g., integrated VHetNet). The terrestrial layer consists of conventional base stations. UAVs, flying aircrafts, and HAPS systems are the main components of aerial networks. UAVs can be used either as an aerial base station or as user equipment. The proposed SEI-VHetNet architecture not only integrates the terrestrial-aerial-satellite networks but also incorporates intelligence and provides a computation and caching platform to enable multi-level edge computing. The distributed computing resources in SEI-VHetNet form the core of the distributed and collaborative computing component of the SE structure. To elaborate on the process in which the SE framework and integrated VHetNet work together, consider the situation of providing communication coverage
for a specific event in a region where there is no terrestrial network coverage. The SE unit first utilizes AI/ML-based techniques to predict the location where the demand for network
capacity and coverage is very high. This can be done using the collected data about node mobility, behavior, service requirements, and used applications. 
Second, by using the prediction outcomes, the SE unit makes an intelligent and automated decision to send and adjust some cross-tier beams, such as UAVs to terrestrial BSs to extend the network coverage to the
event region and accelerate its capacity to support various services for users across the vertical tier during the event time. Through
learning the QoS requirements of nodes' used applications, the SE unit has the capability to select the optimal approach to
backhaul the transmission to the core network through terrestrial, aerial,
or space networks. If some applications
require computational offloading, then the SE unit directs the offloading to the most suitable computational level (e.g., cloud computing, fog
computing, user equipment collaborative computing). Third, the SE unit keeps monitoring and evaluating the network environment by
gathering data about network performance and
measuring user satisfaction levels. Finally, the network
environment evaluation is deployed as feedback to
the SE unit in order to adapt to changes
and make more accurate intelligence, and automated decisions while evolving the network performance. Moreover, throughout this process, coordination management and conflict avoidance of the SE unit resolve any
conflicts that might arise between different components, networks, operators, or tiers. The required computational and communication power to run the SE unit can be
provided through the distributed and collaborative computing component that utilizes the available computational resources of integrated VHetNet. The SE structure is supposed to have several units including one or more local SE units at each vertical tier and one or a limited number of global SE units, which are described as follows.
\begin{itemize}
     \item {\bf{Local SE Unit:}} Local SE units at each vertical tier and one or a limited number of global SE units. Local SE units are embedded at each tier or across a set of interacting networks and operators within the same tier. They can automatically manage and self-allocate the required communications and computational resources to fulfill the constantly changing node demands in the same tier or network. They make their observation of the network environment, user demands, and service requirements, and train and learn the ML models to make automated control and management decisions. The automated decision-making process needs to be performed jointly with continuous collaboration with the global SE unit, to ensure full integration and coordination among all three vertical tiers. This unit can also be treated as a SON in 3GPP standardization which operates in connection with a centralized intelligent unit to make a self-evolving network. 
    \item {\bf{Centralized (Global) SE Unit:}} On the other hand, the global SE unit collects the network control and management data from each tier in order to enable coordination and integration among tiers and support cross-tier control and management decisions collaboratively. The global SE unit has three main components: a) a data collection entity; b) a computation and coordination management entity; c) a distributed and collaborative computing component. The concept of SE capability implements multi-level intelligent network management policies, which can work across different networks, operators, and even ecosystems (e.g., cellular or satellite ecosystems).  SE is supported by advances in ML (e.g., federated learning, online learning, continual learning), the availability of edge and distributed collaborative computing, and also the agility and mobility of network resources.
\end{itemize}
%%%%%
%%%%%%%%%
\begin{figure*}[!t]
    \centering
    \includegraphics[width=130mm,scale=10]{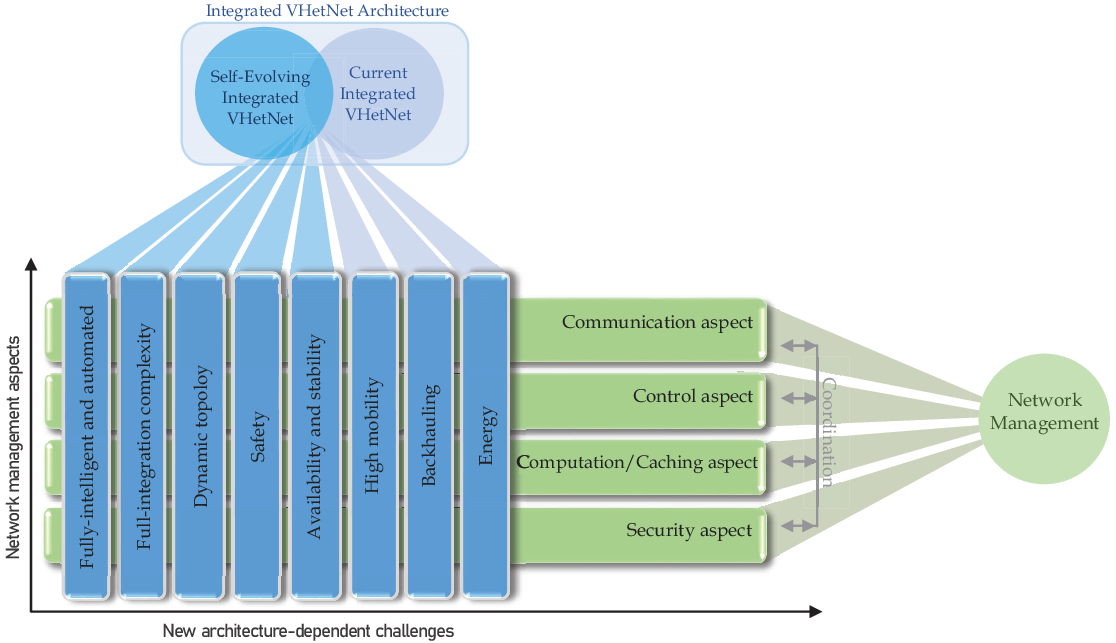}
    \caption{A unified network management framework for SEI-VHetNets.}
    \label{fig3}
\end{figure*}
%%%%%%
%%%%%%%%%%%%%%%%%%%%%%%%%%%
%\subsection{Self-Evolving I-VHetNet}
%The key characteristics of self-evolving networks can identified as follows:
%\begin{itemize}
 %   \item 
%\end{itemize}
\section{Network Management in Integrated VHetNets}
\subsection{Background}
Conventionally, according to the IEEE reference open systems interconnect (OSI) model as a layered architecture~\cite{osi-1}, all network tasks and functions are assigned to specific layers, which basically operate in a hierarchical manner. Each layer provides a set of specific services by introducing a certain set of protocols. Based on the layered architecture, there is no interaction between the layers and they basically operate individually or separately~\cite{osi-2}. The only communication between adjacent layers is to send and receive the processed data. This conventional network management solution will not be suitable for integrated VHetNets. Conventional methods assume that infrastructure changes slowly, over a period of weeks and months~\cite{osi-2}; integrated VHetNets, by contrast, are dynamic with infrastructure that may change in a matter of minutes~\cite{survy20}. Moreover, there is no single model that can cover all possible scenarios. Hence, to design an efficient and adaptive protocol in order to improve the overall network performance along layers involving various network resource management aspects, a novel design is required such that the information between different layers will be exchanged and shared. In such designs, each layer is characterized by a set of specific parameters such that each parameter can be recognized by other layers and they may benefit from it. Furthermore, these designs are mostly formulated as optimization problems, with a set of optimization variables and a number of constraints~\cite{osi-3}. The optimization variables usually refer to the parameters in each layer, and the constraints refer to quality of service (QoS) constraints in different layers~\cite{osi-4}.
\subsection{A Unified Network Management Framework}
\indent In order to jointly tackle the novel architecture-dependent challenges and network management requirements, we propose a unified network management framework as shown in Fig.~\ref{fig3}. In this framework, network management aspects are classified into a set of four network management design aspects, which consist of communication, control, computation/caching, and security. Our aim in proposing this framework is the following:
\begin{itemize}
\item To investigate the network management problems in a more tangible way.
\item To discuss how and why network management in integrated VHetNets is different from conventional approaches. Moreover, we investigate and identify the cooperation and interaction between each VHetNet-specific architecture-dependent challenge and network management requirement, and we show how joint optimization problems can be formulated for each interaction in Section III.
\item To analyze how best to address the resulting highly coupled and complex optimization problems. This opens up the discussion of AI-enabled solutions, which we discuss in more detail in Section IV. 
\end{itemize}
As we can see in Fig.~\ref{fig3}, the unified network management framework consists of two axes: an axis for aspects of network management and an axis for novel architecture-dependent challenges. Each of these is described in detail here.\\
\indent\textbf{Aspects of Network Management Axis:} This axis covers all four network management aspects, namely communication, control, computation/caching, and security. In general, to design a network management policy, these aspects cooperate with each other. Each aspect involves different services and requirements, which are classified here:
\begin{enumerate}
    \item {\em Communication aspects.} This refers to the services and requirements that are mainly involved in transmitting or receiving a signal over the wireless medium. Hence, it includes all the services in the physical layer according to the OSI model: channel modeling, antenna design (including beamforming/precoding), coding/decoding scheme, detection and estimation, rate adaptation, and channel state information (CSI) acquisition.
    \item {\em Control aspects.} This covers any operation or requirement that involves controlling aspects of a wireless communication system. Hence, it includes all  the  services  in  the  data-link  layer  according  to  the  OSI  model:  including interference and collision avoidance policies, admission control, user scheduling, and association policy, network selection control, bandwidth allocation, power control/allocation, and positioning.
    \item {\em Computation/caching aspects.} This covers any operation related to content caching, content offloading to cloud services, and computational tasks. Hence, it includes computation task allocation, content management, and content caching scheduling (cache placement). 
    \item {\em Security aspects.} This refers to the operations, services, and requirements in developing the security techniques for integrated VHetNets. Such techniques may include modeling for eavesdropping and cyber-attack avoidance strategies. Cyber-attacks include control over flight path/hijacking, crashing/landing at will, and access to file system/access to media files. Security aspects basically coordinate encryption protocols and security strategies/technologies across different network layers. Several encryption methods are available at various layers, which can provide end-to-end encryption for the transport, application, and network layers, respectively. To analyze the security problems in greater detail, in this survey, we mainly focus on physical layer security issues and techniques in VHetNets. 
\end{enumerate}

\indent\textbf{Novel Architecture-Dependent Challenges Axis}: This axis shows the core VHetNet architecture-dependent challenges which are basically orthogonal to network management design aspects. These challenges include high mobility, full integration and coordination, dynamic topology, availability and stability, safety, backhauling, and fully intelligent decision-making. Any intersection between an architecture-dependent challenge and an aspect of network management can be interpreted as a new joint optimization problem, which will require a solution, whether analytical or ML-based.
%\subsection{Self-Evolving Capability Requirements}
%%%%%%%%%%%
\subsection{Technical Analysis}
%\indent In the following, 
%%%%%%%%%%%%%%%%%%%%%%%%%%
In this section, we identify and analyze some examples of key architecture-dependent challenges in both current and future integrated VHetNets to demonstrate the requirement for a unified framework. Moreover, according to the unified network management framework, we study the most important joint optimization problems, each as an intersection between a specific architecture-dependent challenge and an aspect of network management.\\
\indent In order to analyze and identify these challenges and their corresponding constraints mathematically, we consider a general VHetNet network model. Let $K$ nodes, fixed or mobile, be randomly dispersed in the terrestrial tier and $M$ aerial nodes hovering in the aerial tier. Moreover, we assume $N$ nodes are dispersed across the satellite/space tier. In the terrestrial tier, the horizontal coordinate of each ground node $k$ is at $w_k(t) = [x_k(t), y_k(t)]$, $k \in K$, at time instant $t$ with $0\leq t \leq T$ where $T$ is the total communication time/period.
In the aerial tier, the nodes may operate at different altitudes with a fixed or variable altitude $H$, above the terrestrial tier. Hence, the time-varying coordinate of each aerial node $m \in M$ at time instant $t$ can be denoted by $ S_m(t)= [q_m(t), H_m(t)]$ where $q_m(t)=[x_m(t),y_m(t)]$. Moreover, since satellite nodes usually fly at fixed altitudes, we can assume the satellite node $n\in N$ has a fixed altitude $\hat{H}_n$ at each time instant $t$ such that its instantaneous location can be denoted by $B_n(t)=[\hat{q}_n(t),\hat{H}_n]$ where $\hat{q}_n(t)=[\hat{x}_n(t),\hat{y}_n(t)]$.
\\

% the most common network management optimization problems are investigated.
\noindent \textit{\underline{Current VHetNet Architecture-Dependent Challenges}.} 
\begin{itemize}
\item {\bf{High Node Mobility Constraints:}}
The high mobility of nodes is the most dominant issue in current VHetNets and it will remain so. The list of mobility or trajectory constraints that one may impose in integrated VHetNets are as follows:
\begin{enumerate}
        \item {\em Maximum speed constraint.} Aerial nodes may fly horizontally or vertically, either with a fixed or variable speed at each time instant. Due to mechanical limitations, they have a limited maximum speed in both directions which needs to be factored into the problem as follows:
        \begin{align}
            &||q^{\prime}_m(t)|| \leq V^{\max}_{XY}, 0 \leq t \leq T, \forall m,\label{mob_const1}\\
            &||H^{\prime}_m(t)|| \leq V^{\max}_Z, 0 \leq t \leq T, \forall m,\label{mob_const2}
        \end{align}
        where $||.||$ denotes the magnitude, and $V^{\max}_{XY}$ and $V^{\max}_{Z}$ are the maximum aerial speeds in horizontal and vertical directions, respectively, in meter/second (m/s). Moreover, $q^{\prime}_m(t)$ and $H^{\prime}_m(t)$ denote the first-order derivative of aerial horizontal and vertical locations, respectively.
        \item {\em Minimum speed constraint.} This constraint is practically valid for rotary-wing aerial nodes capable of remaining stationary at fixed positions, i.e., a minimum zero-speed is feasible. However, for the fixed-wing aerial nodes that must always be moving, a minimum speed constraint must be considered:
\begin{align}\label{mob_const3}
    ||q^{\prime}_m(t)|| \geq V^{\text{min}}_{XY} > 0,\: 0 \leq t \leq T,\:\forall m,
\end{align}
where $V^{\text{min}}_{XY}$ is the minimum speed of the aerial node in the horizontal direction.
        \item {\em Initial and final location constraint.} If any aerial node needs to return to its initial location by the end of a communication period $T$, then the aerial node trajectory needs to satisfy the following constraint: 
        \begin{align}\label{mob_const4}
            q_m(0) = q_m(T),\: H_m(0)=H_m(T), \: \forall m.
        \end{align}
        If an aerial node does not need to return to its initial location after a communication period, this constraint can be relaxed.
    \item {\em Altitude and beamwidth constraint.} If aerial node $m$ has an adjustable antenna beamwidth, the following constraints are imposed on both altitude and beamwidth at each time instant to ensure that all serving nodes (whether terrestrial/aerial/space nodes) are covered by the main lobe of the antenna at any altitude:
    \begin{align}
        &H_m^{\min}(t)\leq H_m(t)\leq H_m^{\max}(t),\:0 \leq t \leq T,\forall m,\label{mob_const5}\\
        &\theta_m^{\min}(t)\leq \theta_m(t)\leq \theta_m^{\max}(t), \:0 \leq t \leq T,\forall m,\label{mob_const6}\\
        &||q_m(t)-w_k(t)||\leq H_m(t)\tan\theta_m(t),\label{mob_const7}\\
        &\quad\quad\qquad 0 \leq t \leq T,\forall k,m.\nonumber
    \end{align}
    
\end{enumerate}
\item {\bf{Backhaul Connectivity:}} Aerial/space nodes can be employed as base stations or relay nodes for backhauling such that uninterrupted connectivity with the core network is maintained. With free-space optical (FSO) communications, several aerial nodes, e.g., HAPS systems, can form a powerful backbone network and enable ultra-low latency backhaul connectivity for aerial and terrestrial network elements. FSO systems are inherently secure and energy efficient while offering huge bandwidths~\cite{R-FSO1}. To extend the coverage, a satellite platform can be employed for the backhaul traffic between the access and the core networks via point-to-point RF/FSO links. To ensure reliable and seamless FSO backhaul connectivity in such three-tier vertical networks, several associated performance parameters need to be taken into account, including the probability of fade, outage probability, and instantaneous end-to-end signal-to-interference-plus-noise ratio (SINR)~\cite{R-FSO2}. For instance, let $\gamma_{m}^{\text{bh}}(t)$, $\hat{\gamma}_n^{\text{bh}}(t)$, be SINR of the links between the aerial node $m$ or space node $n$, and the core network, respectively. Then the following backhauling associated constraints are required to be satisfied:
\begin{align}
    &\gamma_{m}^{\text{bh}}(t)\geq \gamma_{\text{th}},\:0\leq t\leq T, \forall m,\\
    &\hat{\gamma}_n^{\text{bh}}(t)\geq \hat{\gamma}_{\text{th}},\:0\leq t\leq T, \forall n,
\end{align}
where $\gamma_{\text{th}}$ and $\hat{\gamma}_{\text{th}}$ are the defined thresholds ensuring the backhaul connectivity. Such thresholds can also be considered for outage probability and probability of fade~\cite{R-FSO3}. With manual or semi-automated network management in such a complicated vertical network, its capabilities will be limited, the resources will be wasted, and its operational costs will increase significantly. Hence, a fully automated and intelligent network management framework is vital for integrated VHetNets.\\
%\item {\bf{Energy Shortage:}}
\end{itemize}

\textit{\underline{SEI-VHetNet Architecture-Dependent Challenges}.}
\begin{itemize}
    \item {\bf{Dynamic Topology:}} Given that aerial nodes are in general highly mobile and can move in unpredictable ways, vehicular density can fluctuate depending on location and time. Based on their relative speeds, aerial nodes also require sufficient time for any other aerial/terrestrial node to transmit its data. Owing to the dynamic environment and frequent topology changes, maintaining connectivity during the transmission time is difficult. On the other hand, the competitive and autonomous behavior of mobile nodes, and the scalability, in future VHetNets, will require a transformation of routing and handoff strategies into more topology-aware, adaptive, and real-time policies. In particular, there will be a massive number of highly mobile nodes in the near future with differing and competing objectives. To handle such a congested network with massive data traffic along with high node and vertical tier mobility, the following constraints in designing routing and handoff strategies need to be considered:
    \begin{enumerate}
        \item {\em{Real-time mobility-aware routing.}} Depending on the application, the QoS and QoE requirements, and the routing protocol need to respond accordingly to achieve the ultimate goal. In the literature, proposed routing algorithms can be classified into several types, including reactive routing, hybrid routing, position-based routing, security-based routing, cluster-based routing, SDN-based routing, and DTN-based routing. In general, depending on the network model, each routing scheme has its own advantages and disadvantages. However, in integrated VHetNets, two additional aspects need to be taken into account: 1) Due to the highly dynamic nature of the network topology, a number of novel constraints are imposed on routing protocol designs; 2) to achieve self-evolving capabilities, an intelligent and topology-aware routing protocol must be employed. In this paper, we focus on delay constraint routing, energy-efficient routing, and secured routing. The corresponding constraints we consider are (1) the probability of successful packet transmission and (2) the delay of an average packet transmission. 
        \item {\em{Mobility-aware handoff.}} The inherited mobility of aerial nodes, e.g., UAVs, leads to frequent switching from one serving BS to another, i.e., handoff. As such, an additional amount of information is exchanged, which degrades further the transmission links. Even though the energy consumption related to the communication overhead is relatively small compared to the aerial nodes' motion energy, handoff events should be kept to the minimum in order to guarantee network reliability and stability~\cite{R1-handoff}. Hence, one major goal in integrated three-tier networks is to minimize the handoff events. In~\cite{R2-handoff}, the authors investigated the disconnectivity and handoff aware path planning problem for the cargo-UAV, aiming to minimize both its energy consumption and handoff rate. However, this type of offline approach may not be suitable for integrated VHetNets. In contrast to conventional handoff protocols, in integrated VHetNets, handoff strategies need to be designed with respect to the connectivity between the three vertical tiers. For this reason, intelligent and topology-aware strategies are vital to maintaining connectivity and stability. The self-evolving intelligent and automated management approach enables the self-deployment of UAV-BSs handles their fast mobility and handoffs and manages the connections of the hundreds or possibly thousands of users that are served by the UAV-BS. In \cite{R3-handoff}, a speed and direction-based call admission control scheme was developed for a standalone HAPS system with the objective of reducing the handoff call dropping probability as much as possible, as forced termination is less desirable than the blocking of a new call. For this scheme, the system continuously tracked the SIR received from the user equipment (UE)'s serving BS's pilot channel and the next strongest SIR received from the UE's neighboring base stations' pilot signals. It was used to derive the speed and direction of the mobile UE relative to the rest of the UEs.
        %In \cite{}, the authors studied high altitude on-the-move flying wireless access points powered by renewable energy. The access point allocates its}
    \end{enumerate}
    \item {\bf{Full Integration and Coordination:}} 
    In order to achieve full integration and coordination between the three vertical tiers, it is necessary to set a number of constraints on network management. The main constraints for this are (1) non-singular connectivity, (2) carrier aggregation, (3) minimum rate requirement, and (4) delay requirement. Moreover, by introducing novel use-cases and applications, such as digital twin~\cite{twin}, virtual reality (VR), and e-surgery~\cite{vr}, a set of new constraints may also be imposed, which will need to be addressed to ensure full integration and connectivity. For future delay-sensitive applications in SEI-VHetNets, such as VR, the interaction latency constraint across the vertical tiers is expected to be one of the main constraints in optimizing the long-term QoE for users of these applications and services~\cite{e-s}.
    %%%%%%%%%%%%%%%%
    \item {\bf{Availability and Stability:}} The availability of nodes in the communication process and the stability of their links at each tier highly depend on how they are associated with other nodes in the network (e.g., aerial base station). In order to ensure the availability and stability of wireless communication between terrestrial and non-terrestrial nodes, the following set of constraints are expected to be imposed on the network.  
    \begin{enumerate}
        \item \textit{Conventional association constraint.} Considering the mobility of both terrestrial and aerial nodes, in each time slot, to ensure the availability of the communication link, the node association needs to be specified along with the node communication scheduling. Each aerial node (as a base station) may serve a group of other terrestrial/aerial/space nodes; hence, each terrestrial node can be associated with at most one aerial node. This aerial-terrestrial node association imposes a constraint on the network. In general, a binary variable $\alpha_{k,m}[n]$ is defined to indicate that node $k$, as a terrestrial node, is served by aerial node $m$ at time slot $n$ if $\alpha_{k,m}[n]=1$; otherwise,  $\alpha_{k,m}[n]=0$. This imposes the following constraints at each time slot:

    \begin{align}
        &\sum_{k=1}^K \alpha_{k,m}(t)\leq 1,\:0 \leq t \leq T,\forall m,\\
        & \sum_{m=1}^M \alpha_{k,m}(t)\leq 1,\:0 \leq t \leq T,\forall k,\\
        & \alpha_{k,m}(t)\in\{0,1\}, \:0 \leq t \leq T,\forall k,m.
    \end{align}
    Similarly, for the association between aerial nodes and satellite nodes in the space tier, the following association constraints are imposed:
    \begin{align}
        &\sum_{m=1}^M \beta_{m,n}(t)\leq 1,\:0 \leq t \leq T,\forall n,\\
        & \sum_{n=1}^N \beta_{m,n}(t)\leq 1,\:0 \leq t \leq T,\forall m,\\
        & \beta_{m,n}(t)\in\{0,1\}, \:0 \leq t \leq T,\forall m,n,
    \end{align}
    where $\beta_{m,n}(t)$ is a binary variable such that aerial node $m$ is served by satellite node $n$ if $\beta_{m,n}(t)=1$; otherwise, $\beta_{m,n}(t)=0$.
    \item \textit{New association constraints.} The conventional association policies will not be able to address dynamic and collaborative approaches in future networks.  In particular, the nodes in different vertical tiers may not need to be associated with a single server or base station. One promising approach involves multiple nodes transmitting their messages simultaneously, and multiple receivers across the tiers decoding them collaboratively. Hence, these novel association techniques are expected to impose some new constraints on the problem, which should be taken into account. 
    %In general, the collision avoidance constraint is a nonconvex constraint with respect to 3D location/trajectory variables. Moreover, the association constraints are basically integer constraints; hence, making the optimization problem a mixed-integer non-convex problem.
    \item {\em{Energy and power constraints.}} Energy will be consumed by distinct operations of aerial nodes, including flying, communication, computation, and sensing, each with specific power requirements. Due to the limited power budget of aerial nodes, a general constraint may impose on the power management of these nodes. Hence, for an aerial node $m$, $m \in M$ at time instant $t$, with the energy consumption required for flying $p^f_m(t)$, communication $p^t_m(t)$, computation $p^c_m(t)$, and sensing $p^s_m(t)$, its total available power budget is subject to the following constraint: 
    \begin{align}
        0 \leq  p^f_m(t)+& p^t_m(t)+p^c_m(t)+p^s_m(t) \leq P^{\max}_m,\nonumber\\&0 \leq t \leq T,\forall m,
    \end{align}
    where $P^{\max}_m$ is the maximum available power of aerial node $m$ \cite{cog1}. Subsequently, the maximum energy constraint can also be represented as follows: 
    \begin{align}
        &\int_{0}^T p^f_m(t)\text{d}t+\int_{0}^T p^t_m(t)\text{d}t+\int_{0}^T p^c_m(t)\text{d}t\nonumber \\&+\int_{0}^T p^s_m(t)\text{d}t\leq E^{\max}_m, \forall m,
    \end{align}
    where $E^{\max}_m$ is the maximum available energy of aerial node $m$ \cite{cog2}. 
    \end{enumerate}
    \item {\bf{Safety:}} By safety, we mean the safety of humans and the infrastructure accounting for the autonomous movement of nodes.  
    %The safety issue can mainly be addressed by introducing the following constraints, which ensure collision avoidance by aerial/space nodes.
    \begin{enumerate}
    \item {\em{Collision avoidance constraint.}} In case there are multiple nodes in an aerial or space tier, each node must keep a minimum distance from other nodes to avoid a collision. Thus, this constraint can be defined as follows:
        \begin{align}
            &||q_m(t) - q_{m^\prime}(t)||^2 + ||H_m(t) - H_{m^\prime}(t)||^2\geq d^2_{\min},\\
            & ||\hat{q}_n(t) - \hat{q}_{n^\prime}(t)||^2\geq \hat{d}^2_{\min},
            \\& 0 \leq t \leq T, \: \forall m^\prime \neq m,n^\prime\neq n,\nonumber
        \end{align}
        where $d_{\min}$ and $\hat{d}_{\min}$ denote the minimum safe distance between the aerial nodes and satellite nodes, respectively, to ensure collision avoidance in both aerial and space tiers.
        \item {\em{Turning angle constraint.}} This constraint limits the aerial nodes' horizontal turning angles to avoid the phenomenon of so-called chattering. With turning angle constraints, an aerial node cannot arbitrarily adjust its horizontal flight direction, especially for fixed-wing aerial devices. This constraint can be defined as follows:
    \begin{align}
        \phi_{m}(t)\leq& \phi_{m}^{\max} \xrightarrow{} \cos \phi_{m}(t)\leq \cos\phi_m^{\max},\\& \:0 \leq t \leq T,\forall m,\nonumber
    \end{align}
    where $\phi_{m}^{\max}$ is the maximum turning angle of aerial node $m$.
        \end{enumerate}
        %%%%%%
        \item {\bf{Multi-Tier Interference:}} Within each vertical tier, there can be two types of interference, including intra-tier interference in the licensed spectrum, which is the same as inter-cell interference in a cellular network, and inter-operator interference in the unlicensed spectrum. On the other hand, there can be interference across the tiers, which we refer to as multi-tier. Control and management of this interference is a crucial task in enabling the self-evolving capability by supporting full integration and coordination across tiers. To avoid this interference, one approach is that all base stations can be collocated on the same tier and exploited for uplink or downlink connections~\cite{R1-int}. This feature is supported in aerial nodes, such as HAPS, and unlike terrestrial cellular networks, can allow the exchange of information on the interference conditions within the service area between base stations with no signaling overhead. Then, if the total power at any BS is less than or equal to a power outage threshold, the call gets accepted; otherwise, it gets blocked.\\ Moreover, the mobility of nodes across the service area, which can be an integration of two or three tiers, has an impact on how power should be controlled such that node or user admission is optimized.
        %Only a few publications have considered the mobility of UEs in power control of HAPS systems. \cite{} considered a hierarchical system in which a single HAPS and a terrestrial cellular network were jointly deployed, as illustrated in Fig. 
        In these networks, some of the aerial nodes at higher altitudes, such as HAPS, can be used to provide  coverage for base stations at lower tiers, e.g., SMBSs, and the terrestrial cellular towers can be used for macro-cell coverage at a different frequency band, therefore multi-tier or cross-layer interference is avoided~\cite{R2-int}.
        \item {\bf{Fully Intelligent and Automated Decision-Making:}} One fundamental requirement of integrated VHetNets with self-evolving capability is to support intelligent decision-making to provide online, automated, and adaptive network management policies. ML can be effective in learning from experience and detecting changes. Thus, with such knowledge and self-awareness, continuous, intelligent, and automated decision-making can be made to evolve the network. For example, intelligent decisions can be made to inject more communications/computation resources/components, or add services in the network when there are expected demands for extra high data rates or ultra-low latency edge computing~\cite{R1-ML}. Intelligent management services can support satellite, HAPS, UAV, and terrestrial networks' requirement to be self-controlled and self-managed with automated decision-making capabilities. Likewise, satellite global coverage can support the multi-level data collection and computational offloading required for self-evolving functionalities. To enable this automated and evolving capability, a novel architecture with advanced AI/ML-based policies which are discussed in detail in Section V. 
\end{itemize}
%\subsection{General Optimization Problem}
\subsection{Unified Framework: Optimization Problem Analysis}
A general optimization problem can be formulated as follows:
\begin{align}\label{opt1}
    &\min_x f(x)\\&\text{s.t.}\nonumber\\
    & h_i(x) \leq b_i, i = 1, 2,\dots,m,\nonumber\\
    & g_j(x) = c_j , j = 1, 2,\dots, n,\nonumber
\end{align}
where $x$ is the set of optimization variables. The function $f(x)$ is the objective function. The constraint conditions $h_i(x) \leq b_i$ and $g_j(x) = c_j$ is the inequality and equality constraints, respectively. If there is no constraint, the problem is unconstrained. The optimization problem formulated in \eqref{opt1} describes the problem of finding an optimal $x^*$ that minimizes $f(x)$ among all $x$ satisfying the constraints $h_i(x)\leq b_i$ and $g_j(x) = c_j$. Therefore, $x^*$ is called the optimal solution of the problem \eqref{opt1}. \\
\indent Convex optimization is an important class of optimization problems. A standard convex optimization is defined as where $f(x)$ and $h_i(x)$ are convex functions. Compared to problem \eqref{opt1}, the convex problem is such that the objective function and inequality constraint functions must be convex, while the equality constraint functions $g_j(x) = d^T_j x-c_j$ must be affine, where $d^t_j$ is the transposed vector of $d_j$. Convex optimization problems can be solved optimally by many efficient algorithms, such as interior-point methods. If a practical problem can be formulated as a convex optimization problem, the original problem can then be solved. Therefore, many problems can be solved via convex optimization by transforming the original problem into a convex optimization problem. Another class of optimization problems is non-convex optimization, which covers problems with non-convex objective functions or/and non-convex constraint functions~\cite{boyd}. Non-convex optimization problems are usually intractable. The complexity of global optimization methods for non-convex problems may grow exponentially according to the size of the problem. However, some non-convex problems can be transformed into or approximated by convex problems. By solving the resulting convex problems, we can get the optimal solution for the original non-convex problems. Moreover, to overcome the difficulties of solving non-convex problems, some heuristic algorithms can be designed on the basis of convex optimization, such as randomized algorithms where an approximate solution to a non-convex problem is found by drawing candidates from a probability distribution and taking the best candidate as the approximate solution~\cite{opt}. \\ % In addition, for non-convex problems, the compromise is to give up seeking the optimal solution. Instead, we seek a locally optimal solution by combining convex optimization with a local optimization method, where convex optimization can be used for the initialization of local optimization.\\
%\subsection{Unified Framework: Optimization Problem Analysis } 
\indent In the following, we analyze network management optimization problems involving the coordination of any architecture-dependent challenges and network management aspects according to the proposed unified management framework.
\begin{itemize}
\item {\bf{Joint Mobility Model and Communication Design}:} According to the unified network management framework, the intersection between the VHetNet architecture-dependent challenges and the communication aspect leads to a set of joint optimization problems. In these problems the main goal is mainly to jointly optimize the beamforming policy, preceding scheme, adaptation parameters, and aerial mobility design, under specific channel models~\cite{jmcom10}, \cite{omid}. In~\cite{jmcom12}, the direction of 3D beamforming and the trajectory of the aerial node were optimized, subject to the trajectory constraints. In another work~\cite{jmcom13}, the authors proposed a joint optimization problem for statistical precoding, 3D trajectory design, and user scheduling for UAV base stations while maximizing the sum rate of nodes. 
%Moreover, in~\cite{jmcom1}, the beam sizes at the transmitter and receiver.
Also, in \cite{jmcom2}, joint beamforming, user association, and UAV-height control problem was formulated for cellular-connected multi-UAV communications, where the objective was to maximize the minimum achievable rate for UAVs subject to co-existing terrestrial node rate constraints. More recently, in \cite{jmcom3}, authors aimed to maximize the user achievable rate via jointly optimizing UAV trajectory, transmit precoder, and sensing start instant, subject to the sensing frequency and beam pattern gain constraints. Authors in~\cite{jmcom4} also jointly designed the UAV flight trajectory together with the transmit beamforming for optimizing the communication performance while ensuring the sensing requirements.
\item {\bf{Joint Mobility Model and Control Design}}: According to the unified network management framework, the intersection between the control aspect and high-mobility challenge leads to a set of novel optimization problems where the mobility model of aerial nodes need to be jointly optimized along with controlling policies. The mobility model may be defined as a 3D trajectory/ path planning or 3D placement of aerial nodes. Considering a joint 3D path planning and resource allocation optimization problem, the corresponding problem can be represented as \eqref{opt1}, such that constraints $h_i(x)$ and $g_j(t)$, where $i=1,\dots,m$, $j=1,\dots,n$, are divided into two groups including the mobility constraints, i.e., eq.~\eqref{mob_const1}-\eqref{mob_const7}, and the conventional well-known constraints assuring the resource allocation requirements~\cite{Co1}, user scheduling and association requirements~\cite{Co2}, and/or power control requirements~\cite{Co3}. Moreover, the objective function generally is a network performance metric, which can be defined on the basis of the proposed system model. The most popular performance metrics for this type of problem are network total throughput/ sum rate~\cite{Co4}, geometric mean rate~\cite{Co5}, minimum rate (among all terrestrial/aerial nodes)~\cite{Co6}, outage probability~\cite{Co7}, and proportional fairness~\cite{Co8}. Furthermore, in \cite{rout1}-\cite{rout5}, joint 3D path planning and routing optimization problems are formulated in order to find a joint optimal routing policy and 3D mobility model while maximizing/minimizing the aforementioned network performance metrics. Most recently, in \cite{rout6-new}, authors jointly optimized spectrum allocation, power control, co-channel link pairing, and content placement, in order to improve the UAV's energy efficiency. \\
    \indent These types of problems in their general form are non-convex mixed-integer programming due to the non-convexity of objective function and constraints with respect to the 3D location variables.

%\end{enumerate}
\item {\bf{Joint Cache Management and Mobility Model:}}
This problem is mainly formulated as a cache-aided throughput maximization, minimum rate maximization, or average delay minimization problem in current VHetNets by jointly optimizing the trajectory of caching aerial nodes (e.g., UAVs, content caching policy, and cache data access node selection)~\cite{chach1}-\cite{chach11}. The main constraints which are considered in problem formulations are energy availability constraints, QoS requirements, such as delay or outage, and trajectory constraints, (i.e., Eq.~\eqref{mob_const1}-\eqref{mob_const7}). There are also some studies on joint considerations for security and cache placement while optimizing the mobility model of aerial nodes~\cite{chach8}, \cite{chach10}. In most of the recent works~\cite{cache1-new}, \cite{cache2-new}, the formulated optimization problems are NP-hardness. To tackle these problems efficiently, authors decomposed the optimization problem into some sub-problems, including UAV deployment or trajectory problem and content placement problem.
%\begin{enumerate}
\item {\bf{Joint Secure Communication and Control Design:}}
According to the unified network management framework, this problem is basically the intersection of the security aspect which is also coordinated with the communication and control aspects, and the high mobility challenge of integrated VHetNets. In general, jamming and eavesdropping between aerial and ground devices are two major security problems in integrated networks~\cite{sec0}, \cite{sec00}. Towards general optimization problem \eqref{opt1} in physical layer security, the objective function $f(x)$ may be the considered performance metrics, such as secrecy rate/capacity, secrecy outage probability/capacity, and power consumption. In the design of secure beamforming/precoding and 3D path planning of multi-aerial nodes located in the aerial tier, the set of optimization variables $x$ will be the beamformer/precoder and the 3D location of aerial nodes~\cite{sec1}. In~\cite{02} and \cite{03}, optimization problems were formulated to maximize the average secrecy rates of the integrated aerial-terrestrial transmissions, by jointly optimizing the UAV's trajectory, and the transmit power. Most recently, authors in~\cite{sec001}, formulated a joint power allocation and aerial jamming problem to achieve reliable and secure communications for the UAV-enabled NOMA communication system in the presence of a malicious eavesdropper.\\
%In general, we can roughly list several optimization problems which usually appear in physical layer security as follows. 
%%\begin{enumerate}
 %   \item  Integer programming where some or all optimization variables are constrained as integer values. This kind of problem usually arises in the design of secure subcarrier allocation and antenna/node selection.
  %  \item Mixed integer programming where problems have discrete and continuous variables. Such problems are usually dealt with in joint subcarrier allocation, power allocation, or joint antenna/node selection and beamforming.
  %  \item Difference of convex functions programming where the objective function is a subtraction of two convex functions. This feature fits with the definition of secrecy rate/capacity. Therefore, the difference of convex functions programming is widely used for solving the problems of secrecy rate maximization. 
  %  \item Quadratic programming where the objective function has quadratic terms. This problem appears in the designs of secure power allocation and beamforming, such as the typical optimization problem of power minimization.
  %  \item Semi-definite programming which optimizes a linear function of the variables subject to linear equality constraints and a nonnegativity constraint on the variables.
  %  \item Fractional programming which focuses on optimizing a ratio of two nonlinear functions. \end{enumerate}
\indent Specifically, the main idea of secure beamforming is to compute the optimal beamforming vector for achieving some performance metrics of physical layer security by enhancing the signal quality at the destination node and decreasing the signal quality at the eavesdropper while optimizing the 3D path planning of aerial nodes. Due to the special form of logarithmic subtraction in the secrecy rate, the optimization problems of secure beamforming are usually neither convex nor concave in many situations~\cite{sec2}. Most recently, the authors in~\cite{R1-secure} considered the uplink of the full-duplex UAV-aided three-layer space-air-ground communications networks, comprising of terrestrial IoT terminals, a UAV, and an LEO satellite, where eavesdroppers are intercepting the information transmitted. To ensure a secure uplink transmission, a joint UAV deployment and power allocation scheme were developed for maximizing the secrecy rate, subject to the following constraints: 1) UAV's power, 2) the UAV deployment area, and 3) the secrecy rate, which are imposed on the different layers.\\
By contrast, due to the property of logarithmic subtraction in secrecy rate/capacity, security problems are mostly non-convex with respect to beamforming/precoding variables. Moreover, considering the non-convexity introduced by the high mobility and the beamforming objective function of the secrecy rate, the network management problem of designing joint secure beamforming and 3D path planning can only be solved sub-optimally or by numerical methods with high complexity. To mitigate the computational cost of numerical methods, some low-complexity sub-optimal algorithms have been proposed to simplify the beamforming designs~\cite{sec3}. However, to obtain an optimal network management design, low-complexity AI/ML-based algorithms with a high convergence speed can be employed. These are discussed in Section IV, where we introduce and study intelligent network management for integrated VHetNets.
%\item ... 
%\end{enumerate}
\end{itemize}
%%%%%%%%%%%%%
\section{AI/ML-Empowered Solutions}
\subsection{Background}
AI/ML is increasingly employed in communications networks to cope with two main issues inherent in these networks: 1) lack of accurate mathematical model (e.g., unavailability of up-to-date CSI for power control) or when the objective functions/constraints are difficult to be presented in closed-form mathematical expression (e.g., value-of-information or information freshness); 2) intractability of the dynamic mathematical model (e.g., the impact of mobility of coverage or the blockage detection via RF and Img). It has been suggested that ML can be used to reduce the complexity of optimization problems, but usually, the problem should be homogeneous, i.e., continuous or combinatorial (e.g., minimum cut or shortest path). For such cases, the best way to guarantee a reduction in complexity is by turning the algorithm into a machine-learning structure.\\
%\textcolor{red}{2-in any event, this introductory part provides several strong opinions about several broad subjects, without any backups.}
\indent As discussed in the previous section, the network management problems are usually mixed-integer non-linear programs (MINLP), which are generally difficult to solve due to binary variables and non-convex objective functions or constraints. A relaxed version when the binary variables are allowed to take any value between $0$ and $1$ may be considered; however, the relaxed version may still be non-convex due to variables such as trajectory, in the objective function or constraints. To tackle this issue AI/ML-empowered solution methods have shown significant potential and effectiveness. ML has recently emerged as a disruptive technology to bridge the gap between computational complexity and performance in various optimization problems. This trend has encouraged researchers to apply the advances of machine learning to effectively address various problems and challenges in the areas of future networking.\\
\indent By contrast, when we consider an SEI-VHetNet, AI/ML is supposed to also have a novel application in coping with new issues. Specifically, to ensure full integration and coordination between the three tires, using advanced ML methods seems inevitable and necessary.\\
\indent Hence, in the following subsections, we first investigate the applications of AI/ML methods in current VHetNets. Later, we focus on the application of AI/ML in SEI-VHetNets and identify the corresponding core challenges and requirements.    

\begin{figure*}[!t]
    \centering
    \includegraphics[width=160mm,scale=10]{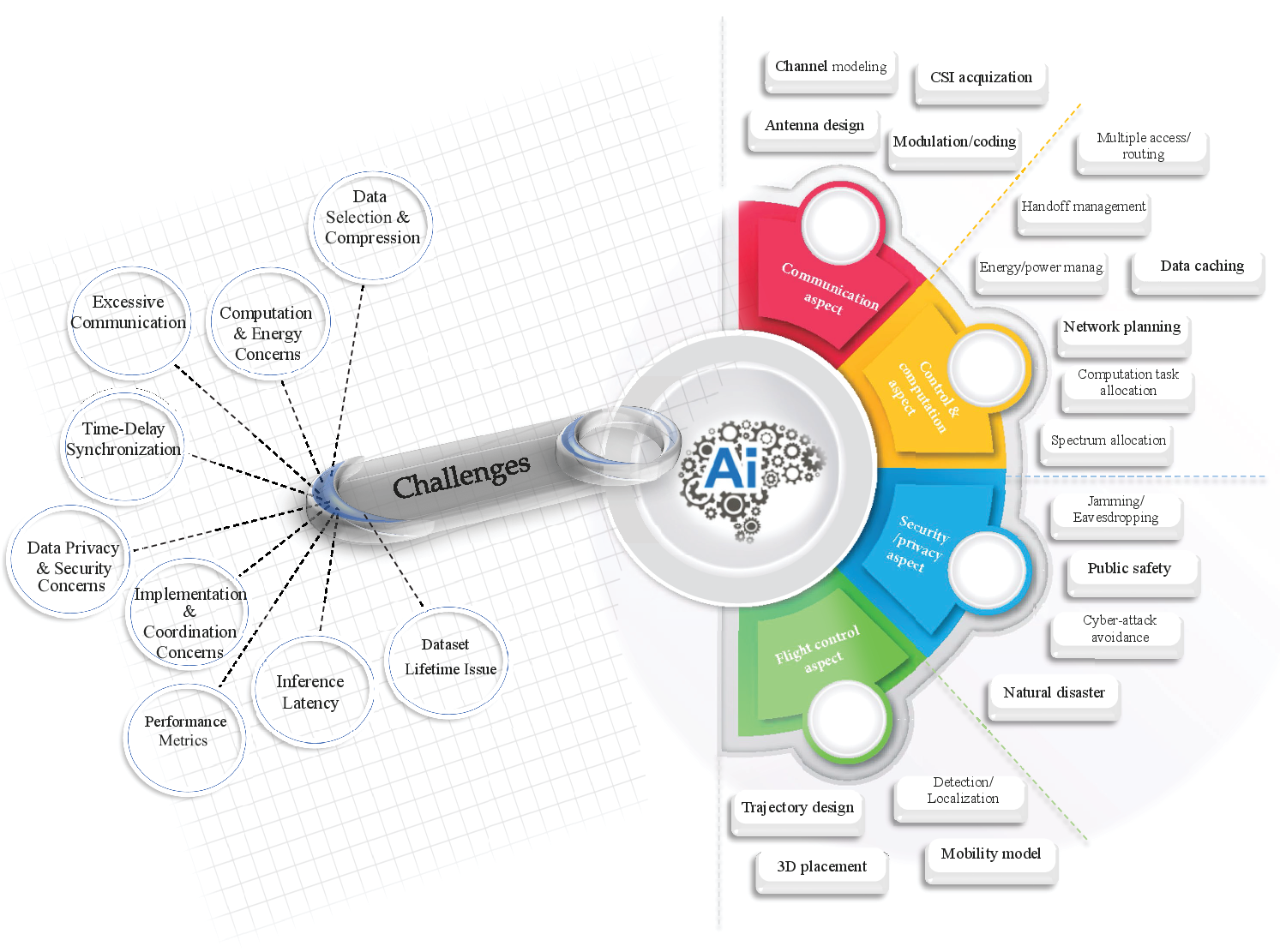} 
    \caption{Applications and challenges of integrating AI/ML solutions in SEI-VHetNets.}
    \label{fig4}
\end{figure*}
%%%%%%%%%%%%%%%%%%%%%%%%%%%
\subsection{Applications of AI/ML Solutions in Current V-HetNets}

%\textcolor{red}{A very succinct introduction. No need to be comprehensive about machine learning and AI, but too brief introduction causes misunderstanding from the reader side.}
The ultimate goal of network management is to intelligently control the network operations without any intervention and still be able to cope with the unknown and challenging requirements of novel services. 
%The aim is to manage the network resources adaptively and real-time in an optimal manner. 
To this end, ML algorithms have been proposed as an efficient approach for addressing the various challenges of IoT ecosystems~\cite{AI4}. In general, ML is based on a pattern recognition framework and its main idea is to exploit correlation among a set of data and/or past good action sequences for adapting to environmental changes without any kind of human intervention. Clearly, the advantage offered by ML in wireless network operations is that it can enable network elements to monitor, learn, and predict various communication-related parameters, such as wireless channel behavior, traffic patterns, user contexts, and device locations~\cite{AI5}.\\
\indent In particular, intelligent network management seems to have the most dominant role in envisioning an integrated VHetNet architecture with self-evolving capability, i.e., SEI-VHetNet. Note that: 1) joint network optimization problems as described in Section IV are highly complex and coupled, so they are computationally prohibitive to be solved optimally; 2) almost all proposed analytical solutions are offline resulting in static network management;
%\textcolor{red}{There should be online ones too!. Also, ML/AI solutions are basically offline in that sense. Do you know any trustable online machine learning solution that learns from scratch in real time?},
3) the highly dynamic nature of integrated VHetNet architecture, which demands a real-time integration and coordination process between the tiers; 4) the life-cycle cost of running a mobile vertical network, which requires eliminating manual configuration of network elements at the time of deployment through dynamic and intelligent optimization and troubleshooting during operations; 5) the need for adaptive and real-time responses to novel and dynamic user service requests and to improve network performance and customer experience~\cite{AI-plc1}. 
\noindent In Fig.~\ref{fig4}, the main applications of AI/ML solutions are highlighted for each network management aspect, along with the expected challenges and issues for SEI-VHetNets. In the following, we discuss some of these applications which are popular both in academia and industry. Moreover, in TABLE~\ref{tab2}, proposed ML solutions for optimizing network management designs while improving various network performance metrics are listed. 
%%%%%%%%%%%%%%%%%%
\begin{table*}[!t]\label{tab2}
\caption{Intelligent Network Management in Current VHetNets}
\begin{center}
\begin{tabular}{L{1cm}|L{3.5cm}|L{3cm}||L{1cm}|L{3.5cm}|L{3cm}}
\toprule %
\textbf{Ref.}&  \textbf{Network Performance Metric and Target}& \textbf{ML Solution Employed}&\textbf{Ref.}&  \textbf{Network Performance Metric and Target}& \textbf{ML Solution Employed}
\\[.3\normalbaselineskip]
\hline 
\cite{AI-plc4} &Throughput and path planning &Multi-agent Q-learning&\cite{AI-nrm1}&Packet transmission probability and connectivity& LR and SVM \\[.3\normalbaselineskip]
\hline
\cite{AI-plc7} &Sum-rate&Q-learning&\cite{AI-nrm3}&Weighted sum-rate and user association& K-means\\[.3\normalbaselineskip]
\hline
\cite{AI-plc4b} &Throughput and 3D placement &double Q-learning&\cite{AI-nrm2}&Prediction accuracy& CRF \\[.3\normalbaselineskip]
\hline
\cite{AI-plc4c} &Task completion time and 3D placement &dueling double deep Q-learning&\cite{AI-nrm4}&Network throughput and mobility control& RL and k-means \\[.3\normalbaselineskip]
\hline
\cite{AI-plc6} &Throughput&GP and NMPC based&\cite{AI-nrm2b}&Communication coverage and mobility control& RL  and  swarm  intelligence \\[.3\normalbaselineskip]
\hline
\cite{AI-plc8}&Communication quality&NN&\cite{AI-nrm6}&QoS and user association& ESN algorithm\\[.3\normalbaselineskip]
\hline
\cite{AI-plc9}  &MOS and QoE&Q-learning&\cite{AI-nrm5}&Traffic  congestion prediction and power consumption&  Gaussian mixture model  
\\[.3\normalbaselineskip]
\hline
\cite{AI-plc11}&Mobility prediction and object profiling&Unsupervised learning&\cite{AI-nrm7}&Traffic congestion prediction and UAV deployment& Gaussian mixture model \\[.3\normalbaselineskip]
\hline
\cite{AI-plc11b} &Throughput and path planning&DDPG&\cite{AI-mac4}&Delay and, routing and MAC protocols& RL-based protocol\\[.3\normalbaselineskip]
\hline
\cite{AI-plc12} &Throughput and path planning&Map compression based&\cite{AI-mac3}&Access accuracy rate and MAC protocol& SVM-based\\[.3\normalbaselineskip]
\hline
\cite{AI-plc14} &QoE&Q-learning&\cite{AI-mac5}&Average throughput and delay, and MAC protocol & Decentralized Q-learning based \\[.3\normalbaselineskip]
\hline
\cite{AI-plc15} &Sum-rate&Distributed learning&\cite{AI-mac5b}&Throughput and, routing and MAC protocol & RL-based \\[.3\normalbaselineskip]
\hline
\cite{AI-plc16} &Throughput&ANN&\cite{AI-mac6}& Energy-efficiency& MAB \\[.3\normalbaselineskip]
\hline
\cite{AI-plc17} &Radio coverage&Q-learning&\cite{AI-mac33}& Throughput and data traffic balance& DL \\[.3\normalbaselineskip]
\hline
\cite{AI-plc18} &QoS&Q-learning&\cite{AI-tab25} & The number of offloaded tasks and resource allocation& Multi-agant DDPG
\\[.3\normalbaselineskip]
\hline
\cite{AI-plc19} &Spectral efficiency&DQN &\cite{AI-tab26}& Throughput and trajectory design& DRL
\\[.3\normalbaselineskip]
\hline
\cite{AI-plc20} &Coordination of multiple UAVs&Q-learning&\cite{AI-tab27} & Total energy consumption and path planning& MDP
\\[.3\normalbaselineskip]
\hline
\cite{AI-plc21}, \cite{AI-plc23}  &Radio coverage&Decentralized DRL &\cite{AI-tab28}& Sum power consumption and resource management & FL
\\[.3\normalbaselineskip]
\hline
\cite{AI-plc22} &QoS&Double Q-learning &\cite{AI-tab29}&QoE and path planning& DDPG
%&Radio coverage&DRL 
\\[.3\normalbaselineskip]
\hline
\cite{AI-plc24} &Energy efficiency&SMGD &\cite{AI-tab30}&Energy consumption and trajectory design& DDPG
\\[.3\normalbaselineskip]
\hline
\bottomrule 
\end{tabular}
\label{tab2}
\end{center}
\end{table*}
%%%%%%%%%%%%%%%%%%%%%%%%%%%%%%%%%
\begin{enumerate}
    \item {\bf{Joint Communication and Fight Control Aspects:}}
\\ For any aerial node across vertical tiers in integrated VHetNets, the 3D path-planning or placement needs to be optimized along with control and network management policies for supporting different network services and requirements. In this context, there are recently quite a number of studies on developing joint problems. For instance, in~\cite{AI-plc4}, joint optimization of 3D placement and power control in multiple UAVs scenarios were studied which aimed to maximize user throughput and satisfy user rate requirements. The corresponding optimization problem relied on a multi-agent Q-learning-based algorithm in order to determine the trajectory design of UAVs acting as agents.
The results showed that the proposed algorithm can converge to an optimal state.
The results also indicated that the proposed approach can increase the throughput up to $15\%$, while accuracy improves as the size of the dataset increases.\\ \indent In~\cite{AI-plc7}, the objective was to optimize the 3D path planning of a UAV base station while at the same time improving sum-rate using Q-learning. The channel between the UAV and the ground nodes was modeled as a log-distance path-loss. Moreover, a standard table-based approach and a neural network (NN) approach were used to find Q-function approximators. The simulation results demonstrated that the online learning capability could enable the UAV base station to find its timely landing location. The authors in~\cite{AI-plc4b} studied the trajectory optimization problem for a UAV under connectivity constraints. They employed a double Q-learning method to solve the problem. Moreover, a UAV 3D placement optimization problem was formulated in~\cite{AI-plc4c}, to minimize the weighted sum of UAV task completion time and expected communication outage time. The authors proposed a dueling double deep Q network with a multi-step learning algorithm to solve this problem.\\
\indent The authors of~\cite{AI-plc6} employed a Gaussian process (GP) was employed to predict the communication clink quality at random UAV 3D locations in an urban environment. The corresponding channel model could then be used to perform optimal UAV path planning, either in an offline or online fashion based on a GP. It was shown that the offline creation of the communication link quality map, with a GP prior to the start of the mission, outperformed the online creation of the map during the mission without scanning.\\
In~\cite{AI-plc6b}, a network of cellular-connected UAVs was studied, where the objective was to find a tradeoff between maximizing energy efficiency and minimizing both wireless latency and the interference imposed by the terrestrial network. To deal with this problem, the authors proposed a deep reinforcement learning algorithm based on echo state network cells.
\\
\indent An aerial relaying network with mobile ground users was studied in~\cite{AI-plc8}, where a hybrid approach of
learning-based measurements and model-based channel prediction was proposed to design UAV 3D path planning while at the same time maximizing the communication performance improvement of networked nodes. The authors assumed several different types of urban environments, including the effects of path-loss, multi-path fading, and shadowing with empirically known distributions. The proposed NN-based 3D path planning showed significant performance in scenarios where there is limited information about the mission area.\\
In~\cite{AI-plc11}, the authors developed an unsupervised online learning technique for joint mobility prediction and object profiling of UAVs, in order to improve control and communication schemes.
The results showed a success rate up to $90\%$ in profiling mobile objects for a reasonable noise level and a small training dataset compared to conventional data-driven methods. \\
\indent The joint problem of online 3D deployment and movement designs of multiple aerial nodes was studied in~\cite{AI-plc9}. The objective was to maximize the sum mean opinion score of ground mobile nodes while at the same time improving the user quality of experience (QoE). The authors first used a genetic algorithm-based k-means method to obtain the initial cell partition of mobile ground nodes. Then, by using a Q-learning-based deployment algorithm, the optimal 3D placement of the UAV was obtained. The results showed that the proposed algorithms improve the performance and accelerate the converging speed after a small number of iterations.
\\Moreover, in~\cite{AI-plc14}, to provide video streaming services to ground nodes, a UAV base station was considered. The objective was to optimize the UAV path planning, such that the QoE of ground nodes is improved. The authors proposed a Q-learning approach to deal with this problem. The performance analysis showed that the proposed Q-learning-based algorithm can efficiently identify the UAV trajectory and improve the QoE of the streaming services on the ground. \\
In~\cite{ AI-plc11b}, to jointly design a 3D  trajectory for a network of UAVs while at the same time maximizing its throughput, a markov decision process (MDP) problem was formulated. The authors used a deep
deterministic policy gradient (DDPG) algorithm to deal with this problem. The actions taken by the UAV are related to adjusting both the 3D location and the transmission control.\\
\indent The authors in~\cite{AI-plc12} considered a UAV-enabled IoT network for data harvesting. In the uplink scenario, a joint UAV trajectory and node scheduling design problem was formulated to minimize the estimation error of the channel model parameters and to maximize the data traffic between the UAV and static ground  nodes. To deal with this problem, an iterative path planning algorithm was proposed along with dynamic programming techniques. The results showed the benefits of the proposed learning method, proving that the algorithm can converge to at least a locally optimal solution.\\
Most recently, authors in~\cite{AI-tab26}, proposed an energy-efficient DRL-based algorithm for UAV trajectory design and resource allocation. Simulation results showed that the proposed learning-based approach can achieve better performance compared to the benchmark schemes in terms of the total throughput and fairness among the users.

%%%%%%%%%%%%
%%%%%%%%%%%%%%%%%%%%%%%%%%%
\item {\bf{Control and Computation Aspects:}} To meet the requirements of network resource management in current VHetNets and solve corresponding optimization problems efficiently, ML techniques have recently been adopted.
In~\cite{AI-nrm1}, to efficiently predict the success and failure rates in an aerial network, two learning algorithms based on linear regression (LR) and support vector machine (SVM) were employed. The authors showed that the proposed learning methods can train aerial nodes to determine their connectivity with neighboring nodes. Simulation results revealed that the packet transmission probability can be exactly predicted by $0.0000531$ in root mean square error measurement.
%Through timely training, multi-UAV networks can be optimally deployed and accurate clustering can increase the reliability of wireless transmissions.
\\ \indent To predict cell quality for UAVs connected to a cellular network, learning techniques were used in~\cite{AI-nrm2}. A new conditional random field (CRF) position was adopted for predicting the optimal serving cell. In fact, the idea was to exploit the spatial correlation that naturally exists in the nearby positions of aerial channels. The performance analysis under real 3GPP LTE simulation parameters showed that the proposed approach provides better accuracy while improving the overall performance.\\
\indent To decrease the number of unnecessary handoffs in aerial networks, the authors in~\cite{AI-nrm4} proposed an RL algorithm. For mobility control, they also applied a k-means algorithm. The simulation results demonstrated that the hybrid approach of using RL and k-means algorithms can significantly reduce the unnecessary handoffs while at the same time increasing the network throughput.\\
Considering a network of multiple HAPS systems, the authors in~\cite{AI-nrm2b} aimed to improve communication coverage by using learning-based algorithms. For this aim, RL and swarm intelligence was employed. It was shown that the RL algorithm obtains higher peak coverage rates with the cost of slower convergence while the swarm intelligence-based
the approach achieves lower coverage peaks and improved coverage stability and convergence.\\
\indent Aerial devices with device-to-device (D2D) communications can be used to support connectivity in disaster areas. In~\cite{AI-nrm3}, such a scenario was studied. The authors considered multiple UAV base stations in order to serve ground nodes via D2D connections, such that the objective was to maximize the weighted sum-rate. They proposed a k-means-based algorithm to solve this user association problem. The performance analysis demonstrated that the proposed learning algorithm has a better performance with lower computational complexity.\\ 
\indent In~\cite{AI-nrm6}, the authors considered a UAV-assisted LTE-U network, where UAVs were able to access both the licensed and unlicensed frequency bands. Their main objective was to maximize the QoS while at the same time meeting the delay requirements, by designing an efficient network resource allocation policy. To fulfill this, an echo state network (ESN) algorithm was adopted with leaky integrator neurons. The performance evaluation showed that the proposed learning approach achieved up to $20\%$ QoS performance gains for the users which were associated with UAVs.\\
\indent Moreover, a UAV-assisted cellular network was considered in~\cite{AI-nrm5} such that it consisted of multiple UAVs, acting as aerial base stations, and a set of ground base stations. The authors aimed to predict traffic congestion in order to use UAVs as temporary base stations with minimum requirements regarding communication and mobility powers. They proposed a Gaussian mixture model based on a weighted expectation maximization algorithm. The simulation results demonstrated that the proposed learning algorithm can significantly decrease the requirements for the downlink communication power and the mobility power as compared to scenarios without using learning techniques.
\\ To support on-demand services for ground cellular users, authors in~\cite{AI-nrm7} proposed a learning technique, which enables a predictive, efficient deployment of aerial base stations. More specifically, the proposed learning framework was based on a Gaussian mixture model and also a weighted expectation maximization algorithm was introduced to predict the potential network congestion. The numerical results showed that the proposed strategy could effectively improve the collaboration between the aerial clusters.
\\ Most recently, authors in~\cite{AI-tab30}, studied an aerial mobile edge computing architecture, by taking advantage of the UAVs to serve as the flying platform. They aimed to minimize the energy consumption of all the ground users by optimizing the UAVs' path planning, user associations, and resource allocation. In order to solve the multi-UAVs' path planning problem, a convex optimization-based trajectory control algorithm was first proposed. Then, in order to perform fast decisions, a DRL-based algorithm including a matching algorithm was proposed. The results showed that these learning-based algorithms have considerable performance.
%%%%%%%
%%%%%%%%
\item {\bf{Joint Communication, Control, and Security Aspects:}} In recent years, several ML approaches have been adopted to traditional  medium access control (MAC), routing protocols, and security schemes, to improve their support of the highly dynamic nature of current aerial communication networks, enabling intelligent secure network coordination and user scheduling, and routing via dynamic path selection. Along these lines, the authors in~\cite{AI-mac4}, studied a network of multiple drones, which were equipped with GPS and an identical antenna array. To find the routing path with the minimum delay, they adopted an RL-based protocol where all drones were able to exchange their channel status information. The proposed learning-based approach enabled self-configuration and intelligent operation by establishing connectivity through the optimal routing path. The performance evaluation demonstrated that the proposed routing strategy can enhance the network performance over existing ones and also can ensure successful transmission with lower delay.\\
 In order to improve security by detecting the attackers, the authors in~\cite{AI-secu1}, was developed a new ML model to classify the spoofed and authentic signals received by UAVs. In their proposed methodology, several ML algorithms along with K-learning models were deployed in order to select a suitable classification algorithm. To achieve the desired task, they used GPS signal characteristics as features. Although the results showed an enhanced security performance, the adaptivity to highly dynamic topology in an integrated VHetNet still can not be guaranteed. 
 \\ An SVM-based MAC scheme was proposed without demodulation in~\cite{AI-mac3}. The proposed approach included an ML classifier in it and its performance was evaluated using two typical MAC protocols, including time division multiple access (TDMA) and carrier sensing multiple access with collision avoidance (CSMA/CA). The simulation results showed an access accuracy rate up to $90\%$ while for the case without ML, it was below $60\%$.\\  \indent Furthermore, an adaptive MAC scheme was proposed in~\cite{AI-mac5} where a network with multiple aerial nodes was considered. A decentralized Q-learning-based MAC scheme, based on a practical byzantine fault tolerance procedure, was used to determine the MAC protocol, whether TDMA or CSMA/CA, according to the aerial nodes' current situation. According to this approach, each aerial node was able to evaluate its performance and select the best MAC protocol. The performance evaluation showed that the performance of the proposed scheme with respect to the average network throughput, delay, and packet retransmission ratio improved over conventional MAC protocols.\\ \indent
The authors in~\cite{AI-mac5b}, introduced a combination of a novel position prediction-based directional MAC protocol and a self-learning routing protocol based on RL routing protocol, as an adaptive hybrid communication protocol. The simulation results showed that the proposed MAC protocol resolves the directional deafness issue, which occurs when the transmitter fails to communicate with the receiver due to having the receiver's antenna oriented in a different direction. Moreover, the proposed RL-based routing strategy provided a self-evolving and more effective routing scheme, appropriate for autonomous FANETs.\\
 A multi-armed bandit (MAB) technique was adopted in~\cite{AI-mac6}, in order to determine the energy-efficient and processing data offload paths between a source and target UAV. The authors considered a decentralized large-scale edge UAV swarm and their performance analysis showed that the proposed MAB technique could enable energy-efficient and processor-friendly path solutions compared to other solutions. \\\indent In~\cite{AI-mac33}, an integrated vertical network consisting of UAVs, low earth orbit (LEO), medium earth orbit (MEO), and geostationary earth orbit (GEO) satellites were studied. 
%In this scenario, the shortest path routing resulted in higher priority to MEO satellite routing, which, however, did not guarantee improved performance due to periodic traffic bursts. 
The authors employed a DL technique to ensure data traffic balance. More specifically, the main idea was to transfer some portion of traffic to GEO satellites with a reduced computational cost if distributed periodical training was used. The extensive simulation results demonstrated that the proposed DL approach improved overall network throughput while at the same time reducing packet loss rate compared to other routing schemes.
\\
Most recently, authors in~\cite{AI-mac34-new}, developed a smart UAV that has the ability to detect a dynamic intruder target. The problem of optimizing the search path of the UAV was formulated as an RL problem, where the objective is to maximize the search benefits. The proposed RL schemes enable the smart UAV that is able to learn the behavior of the intruder target and detect it with high probability. Simulation results showed that their proposed algorithms outperform conventional approaches.
\end{enumerate}
%\subsection{Potential Approaches for Integrating AI/ML in Self-Evolving I-VHetNets}
\subsection{What is Different in SEI-VHetNets?}
To develop self-evolving capability, intelligent decision-making strategies and automated operations are needed. In integrated VHetNets due to the unique three-tier network architecture and corresponding challenges, discussed in the previous section, the idea of enabling self-evolving capability is even more complex, requiring novel AL/ML approaches. Hence, the most vital requirement of SEI-VHetNets is the capability of being fully intelligent and automated in almost all aspects of the network. The integration of AI/ML techniques in wireless networks can generally leverage intelligence for addressing various issues \cite{survy20}. This combination of AI/ML and integrated VHetNet architecture appears to be strongly correlated in different disciplines and applications and throughout the network, promising unprecedented performance gains and reduced complexity~\cite{AI1}. The self-evolving framework enables integrated VHetNets to automatically
manage and self-allocate the communications and
computational resources required to fulfill the constantly changing user
demands along the vertical tiers.
%Hence, in self-evolving I-VHetNets, the utilization of some new ML tools are inevitable.
However, most of the currently proposed ML solutions often suffer from a lack of convergence as the number of users/samples increases, as well as sample inefficiency due to sparse rewards and high variance in their optimization algorithms. Before effective solutions can be considered, a deep understanding of the challenges and issues in the integration of AI/ML in SEI-VHetNets is absolutely necessary. In the following section, we identify such challenges and issues.

\section{Open Issues and Future Research Directions}
The integration of AI/ML solutions in SEI-VHetNets is expected to impose some fundamental challenges and issues which need to be addressed carefully. As also highlighted in Fig.~\ref{fig4}, in the following, we introduce the core challenges that are the main bottlenecks impeding the large-scale deployment of AI/ML solutions in SEI-VHetNets. 
\subsection{Core Challenges of Integrating AI/ML Solutions in SEI-VHetNets}
\begin{itemize}
    \item {\bf{Data Selection and Compression:}} Most of the samples in the datasets are redundant, carrying almost no additional information for the learning task. A fundamental open question in this learning structure is how to characterize and algorithmically identify a small set of critical samples without losing the full integrity and coordination of SEI-VHetNets. For instance, in the proposed intelligent framework for SEI-VHetNet in Fig.~\ref{fig5}, it may be enough to share small representative datasets among local SE agents instead of sharing the original large ones, which would lead to a significant reduction in power consumption and required communication bandwidth. Hence, there exists a challenge in how to efficiently select the best small critical samples in each dataset~\cite{ch-1}.\\
   \indent Furthermore, large datasets are generally time and energy-consuming in model training processes. This inefficiency can be a big concern in SEI-VHetNets since any delay in decision-making or shortage in the energy budget can lead to a loss of network integrity and coordination thereby failing in satisfying self-evolving capability. Hence, whether or not large datasets can be compressed, and how efficient and accurate this compression can be, also need to be well investigated.  Moreover, a general framework is needed for function approximation, where choosing the samples that best describe the function is done in conjunction with learning the function itself.
    \item {\bf{Computation and Energy Concerns:}}
    The energy consumption of aerial nodes at higher tiers is mainly due to their hovering and transmitting/receiving information, which makes it important to design energy-efficient ML solutions to deal with network optimizations~\cite{ch-4}.\\ Moreover, to outsource the computational needs and minimize the costs of ML solutions, the use of cloud-based computing seems inevitable. Also, with the distributed computational resources in SEI-VHetNet, implementing distributed ML algorithms will be
practical and feasible.
    Local SE units, whether implemented on devices/users (distributed) or in a data center (centralized), have to deal with computationally intensive data analysis requiring considerable energy. This is due to the massive number of samples received from different resources for various network management purposes/services, and for this reason, the local datasets are expected to be uncommonly large and complicated. Hence, cloud-based distributed computation units not only require high-performance systems, but also a strategy to scale up computation size with only modestly increased energy consumption. Such balanced computation scaling for deep neural networks is critical to minimize their energy overhead. Furthermore, the parallel use of many small, energy-efficient computation nodes at each tier to accommodate large computations, has its own challenges, such as how the small computation nodes should be implemented since aerial nodes (e.g., UAVs) are mostly moving and may not be available to cooperate with other computation units. Consequently, the parallel slowdown issue of computational parallelization is expected to be intensified and more difficult to deal with in SEI-VHetNets due to the availability concerns and highly dynamic nature of the vertical architecture.
    \item {\bf{Excessive Communication:}}
    Since distributed learning approaches involve a massive number of devices participating during model training, communication is a critical bottleneck for integrating distributed ML methods as part of local SE units in SEI-VHetNets. To improve communication efficiency and make the distributed ML framework suitable for SEI-VHetNets with massive, heterogeneous devices and networks, it is necessary to develop a communication-efficient method, which can greatly reduce the number of gradients exchanged between the devices and the cloud instead of all gradients information. In order to further reduce communication overhead in these settings, two key aspects need to be considered~\cite{ch-5}: 1) reducing the total number of communication rounds, and 2) reducing the number of gradients in each communication round.
    \item {\bf{Data Privacy and Security Concerns:}}
    Since SEI-VHetNets can provide ubiquitous services across a wide geographic area, the computing and communication capabilities of each device in any tier may vary due to changes in hardware (CPU, GPU), network connectivity (4G, 5G, 6G, WiFi), and energy (battery level). Obviously, system heterogeneity between devices will present some confusion and faults for the learning models and subsequently SEI-VHetNets. For instance, there may be unreliable devices in distributed learning approaches, such as federated learning, which may cause the Byzantine failure of the system~\cite{byzan}. Similarly, hackers may launch active learning-based attacks (like poisoning attacks and backdoor attacks) on heterogeneous devices and cause errors in the learning system. The security vulnerabilities of these distributed learning systems greatly exacerbate challenges, such as mitigating attacks, tolerance, and faults~\cite{ch-6}. Therefore, developing secure and robust learning methods must: 1) defend against malicious attacks, 2) tolerate heterogeneous hardware, 3) achieve robust aggregation algorithms, 4) support large-scale spectrum trading in an untrusted and nontransparent trading environment, and 5) provide decentralized spectrum trading to efficiently manage the trading among the operators and base station in diff rent tiers. Moreover, although distributed learning techniques protect the privacy of each device by sharing model updates
(e.g., gradients information) instead of raw data, from a privacy perspective, concerns regarding/about disclosing data (for example during the interaction between the device and the cloud at each local SE unit) may still exist. For instance, hackers will launch membership inference or gradient leakage attacks to steal local training data from the devices~\cite{ch-6b}.\\
     Recently, blockchain technology is considered a possible solution to address the above challenges because of its advantages of decentralization, anonymity, and trust~\cite{R1-block}. Blockchain is a decentralized ledger-based storage method, which provides a unique tool for secure transactions in a distributed fashion without trusted agents~\cite{R2-block}. Since the SE framework has distributed structure, blockchain technology can well suit this framework supporting high security and privacy for integrated VHetNets. Moreover, encryption techniques can be employed to accomplish security and tackle some of the aforementioned challenges in SEI-VHetNets. The encryption process can be executed by every node in any tier depending on the data type that needs encryption~\cite{R1-encry}. For instance, flight control or communication data encryption can be processed in the flight control mechanism, and transmission data encrypting for the mission can be processed in the mission servers~\cite{R2-encry}. Hence, encryption techniques can also have the potential to be adopted into the SE framework and utilized as a proficient tool for secure communication and classification in highly dynamic environments.
    \item {\bf{Implementation and Coordination Concerns:}} Implementing the ML solution methods at each vertical tier involve two main aspects which are needed to be taken into consideration simultaneously including distributed/centralized implementation and coordination between the ML units. ML model training can be implemented either on-device (distributed~\cite{ch-7}) or partially/totally on a data center (centralized~\cite{ch-8}) at each tier. One challenge with implementation has to do with the nodes selected for distributed implementation, since there may be a massive number of nodes at each tier. Consequently, this raises a question: Is it a good idea to have dedicated nodes for this purpose in such a highly dynamic vertical architecture, or should we have non-dedicated nodes meaning that all nodes need to be provided with an ML training structure? Another challenge has to do with how these ML units should be implemented at each tier such that the online and real-time coordination with other ML units at other tiers is guaranteed in order to make network management decisions (e.g., routing) that involve information from all vertical tiers.
    \item {\bf{Performance Metrics:}} The impact of the architecture-dependent challenges on overall network performance and most importantly on sustainability in terms of full integration and coordination in VHetNets raise an important question: Are the current performance metrics (e.g., accuracy rate~\cite{ch-9}) to evaluate the effectiveness of ML solutions sufficient or do we need to come up with new or modified metrics for evaluating the efficiency of data compression and selection?
%    \item {\bf{Efficient Model Training:}}
    \item {\bf{Time-Delay Synchronization:}} If tasks in one tier are connected or depend on other tasks in any other tier, time-delayed decisions need to be synchronized. Otherwise, it is not going to be real-time decision-making that maintains the full integration and coordination in SEI-VHetNets.
    \item {\bf{Inference Latency:}} In large-scale wireless networks with several connected communication platforms (i.e., integrated VHetNets), where there can be a massive number of nodes, inference latency~\cite{ch-10} can be a big concern if the objective is to make quick decisions to sustain full-integration and cooperation. To handle this problem, the size of deep neural networks needs to be reduced. Hence, designing a quantization-aware training structure~\cite{ch-11} in local SE units may be critical to improve the total throughput by taking advantage of high throughput integer instructions.
    \item {\bf{Dataset Lifetime Issue:}} To have a self-evolving network architecture, the ML datasets of all vertical tiers need to be updated at some point, otherwise decisions will be less efficient and will not ensure the coordination between local SE units; hence, there is a challenge of having a dynamic training where the lifetimes of the datasets are taken into account and updated accordingly over time~\cite{ch-12}.
\end{itemize}
\subsection{Deployment of Advanced ML Solutions}
 To tackle the aforementioned challenges and difficulties of integrating ML solutions in SEI-VHetNets in order to enable self-evolving capability and improve the overall network performance, some propitious learning concepts have recently gained attention due to their remarkable effectiveness in solving complex problems~\cite{AI6}, \cite{AI7}. In the following, we briefly identify and study the main characteristics of these advanced learning approaches and techniques which can meet the requirements of SEI-VHetNets.
 \begin{itemize}
     \item {\bf{Personalized Federated Learning:}}
The federated learning (FL) architecture in its basic form consists of a curator or server that sits at its center and coordinates the training activities. Clients or users are mainly edge devices, which can run into millions in number. These devices communicate at least twice with the server per training iteration. To start with, they each receive the current global model's weights from the server, and train it on each of their local data to generate updated parameters which are then uploaded back to the server for aggregation~\cite{AI-FL1}, \cite{AI-FL2}. This communication cycle persists until a pre-set epoch number or an accuracy condition is reached. Hence, the procedure of FL-based architecture can be divided into three phases: initialization, training, and aggregation. Each phase involves some tasks and operations, which are summarized as follows:
\begin{itemize}
\item \textit{Initialization phase:} In this phase, some actions or operations are required for each user, such as checking the service demands and connection conditions. After this, each user makes a decision on whether to join the closest cloud server or not. Upon joining the cloud, each registered user receives an initialized and pre-defined global model in order to train this model. The global model is shared and transmitted by the cloud server.
\item \textit{Training phase:} In this phase, each user employs its local dataset to update the received global model in an iterative manner. The objective is basically to minimize a loss function at each iteration. After each update, each user transmits the updated training model to the cloud.
    \item \textit{Aggregation phase:} In this phase, all the uploaded updated training models are received and collected by the cloud server. Based on these collected models, the cloud derives a new global model which will be transmitted to a random set of users again. This iterative process continues until the algorithm converges or reaches a halt benchmark.
    \end{itemize}
    
\indent Considering the characteristics of FL and how it operates, in SEI-VHetNets, FL has the potential to play a key role. However, conventional FL schemes only develop a common output for all the users, and, therefore, it does not adapt the model to each user. This is an important missing feature, especially given the heterogeneity of the underlying data distribution for various users in SEI-VHetNets. Hence, a personalized variant of FL seems to be a better solution. This personalization keeps all the benefits of the federated learning architecture, and, by structure, leads to a more personalized model for each user across vertical tiers~\cite{R1-person}. By employing personalized FL, each vertical tier can be enabled to make local decisions on its local network management aspects while transmitting its local ML model to a centralized unit in order to develop and train a global model which is a necessity to ensure full integration and coordination between the tiers.
\item {\bf{Federated Meta Learning:}}
Meta-learning, also known as ``learning to learn,'' refers to ML models that can learn new skills or adapt to new environments rapidly with a few training examples. There are three common approaches: 1) learn an efficient distance metric (metric-based), 2) use a (recurrent) network with external or internal memory (model-based), 3) optimize the model parameters explicitly for fast learning (optimization-based)~\cite{AI-M1}, \cite{R1-meta}. Meta-learning has been widely applied in the field of multi-task Reinforcement Learning. In meta-learning, a meta-model is obtained through a large number of pre-trainings and is able to adapt quickly to novel tasks in a test. Meta-learning opens up a new perspective in multi-task RL. It pursues a compromise in every task. Meta reinforcement learning refers to the problem of learning policy that can adapt quickly to novel tasks by using prior experience on different but related tasks~\cite{AI-M2}. Most recently, in~\cite{HL-2}, a novel and flexible approach was proposed to meta-learning for learning-to-learn from only a few examples. \\ In applications of federated learning in SEI-VHetNets, it is expected that the statistical and systematic challenges in collaboratively training machine learning models across the distributed vertical networks of mobile users are the bottlenecks. Moreover, due to the highly dynamic nature and the massive number of nodes and connections, rapid decision-making strategies are needed to maintain connectivity and coordination between tiers. Otherwise, full integration will fail. To resolve these issues, meta-learning is a natural choice to be adapted into the federated meta-learning framework, where a parameterized algorithm (or meta-learner) is shared, instead of a global model. Hence, meta-learning can improve the effectiveness and convergence speed of ML solutions to satisfy the requirements of SEI-VHetNets.
\item {\bf{Federated Transfer Learning:}}
Transfer learning is the idea of overcoming isolated learning paradigms and using knowledge acquired for one task to solve related ones. Specifically, transfer learning is the ability of a system to recognize and apply knowledge and skills learned in previous tasks to novel tasks. According to this definition, transfer learning aims to extract knowledge from one or more source tasks and apply this knowledge to a target task. Unlike multi-task learning, where the source and target tasks are learned simultaneously, transfer learning focuses on a target task. The roles of the source and target tasks are no longer symmetric in transfer learning~\cite{AI-T1}, \cite{R1-transfer}.\\
\indent In SEI-VHetNets, it is expected that many tasks will need to be solved simultaneously or not; hence, employing the knowledge obtained for similar tasks across the vertical networks or operators can accelerate the decision-making process and also improve precision and convergence in other networks or operators. An integration of transfer learning into FL systems can provide a unique feature to allow knowledge to be transferred across vertical tiers that do not have many overlapping features and users. Therefore, the performance of self-evolving capabilities is expected to improve significantly by using federated transfer learning methods.
\item {\bf{Hierarchical Federated Learning:}}
One of the major issues with current learning methods, such as deep reinforcement learning, is the demand for massive amounts of sampled data. In conventional centralized FL, users transmit their computation results to the parameter server for aggregation at each iteration. However, in large-scale networks, such as integrated VHetNets, this centralized framework may result in high communication latency and thus increases the convergence time. Hence, this poses several challenges in SEI-VHetNets, which were discussed in the previous section. Most recently, to tackle these issues, hierarchical federated learning, as a sample efficient learning approach, is introduced in which multiple servers are employed in parallel to reduce the communication latency~\cite{R1-haira}. Recent studies reveal that hierarchical learning is robust to hyperparameters and can speed up the learning process compared to conventional approaches. In~\cite{HL-1}, due to the non-convex and combinatorial structure of the maximization problem, the authors developed a deep reinforcement learning approach that adapts the beamforming and relaying strategies dynamically. In particular, a novel optimization-driven hierarchical deep deterministic policy gradient was employed and shown to have superior performance. Moreover, in~\cite{R1-haira}, the authors show that a hierarchical federated learning solution can significantly
reduce communication latency without sacrificing the model
accuracy in heterogeneous cellular networks.
\item {\bf{Continual Active Learning:}} 
Continual active learning is an effective approach to mitigate the effects of wireless network degradation on the training data and can help to adapt models to the changing environment by training on a continuous data stream~\cite{R1-active}. This learning algorithm can determine the training samples with the highest quality of information that has a significant effect on the model construction and its accuracy. The high-quality subset of the continuous training samples thus obtained can be provided with a higher level of protection in the wireless channel by means of increased allocation of network resources and priorities. Continual active earning has been widely considered a promising solution in many modern machine learning applications where obtaining a new labeled training sample is expensive and complex due to the highly dynamic environment. The main components of continual active learning include parameterized model, a measure of the model's uncertainty, and an acquisition function that decides based on the model's uncertainty for the next sample to be labeled on a continuous data stream~\cite{AL-1}.\\ 
%This approach has several challenges, including a lack of scalability to high-dimensional data, a lack of theoretical guarantees, and a lack of formal uncertainty measures. More importantly, the acquisition function is usually greedy in the sense that it sequentially finds new samples to be labeled one by one. Consequently, the resulting sub-sampled dataset may not necessarily be a small representative dataset due to the greedy nature of active learning.\\
\indent In SEI-VHetNets, since it is expected that the datasets will be very large due to the massive number of nodes and connections along all three tiers, and the environment will be highly dynamic and changing, the lack of an efficient continuous labeling procedure will cause delays in decision-making. Hence, to decrease the delay in training ML models in the self-evolving framework and resolve the time-delay synchronization issue, employing a continual active learning algorithm seems to be necessary.
\item{\bf{Quantization-Aware Model Training:}}
To resolve the inference latency issue, which is expected to be more significant in SEI-VHetNets, a quantization-aware ML model training algorithm is a promising solution. Quantization can lead to significant memory and
processing power reductions in local SE units in the self-evolving framework. Quantization-aware training is important to achieve higher accuracy during inference, as it models quantization errors during training to match quantization effects during the inference. Quantization awareness is performed by introducing so-called fake quantizations, which means that the quantization is directly followed by a dequantization. Specifically, quantization has a distinct stage of data conversion from floating-point into integer-point numbers. In general, the process of quantization is associated with the reduction of the matrix dimension via limited precision of the numbers. However, the training and inference stages of deep learning neural networks are limited by the space of the memory and a variety of factors, including programming complexity and even the reliability of the system. The process of quantization has become increasingly popular due to the significant impact on performance and minimal accuracy loss~\cite{Q-1}. Furthermore, to ensure that local SE decisions are synchronized with minimum time delay, dynamic models that are trained online will be indispensable. That is, data is continually entering the local SE units at each tier and it is incorporated into the model through continuous updates. This approach also takes the life of samples and observations in datasets into account, thereby reducing computation and energy consumption in local SE datasets.
\item{\bf{Data-Driven Proactive Decision-Making:}}
SEI-VHetNets will require network management optimization to be service-oriented and user-oriented as well as proactive rather than passive. This will open up a new research direction on how terrestrial/aerial/satellite user demands,
contexts, and experiences can affect SEI-VHetNet network optimization problems which are modified by the new architecture-dependent challenges. The network optimization will be boosted if it can understand user behavior and spatial-temporal traffic patterns through application data along the vertical tiers. In this case, the optimization needs the status information of concrete entities in social space (e.g., UEs) or virtual entities in cyberspace (e.g., software). Such user-centric meta-information is generally called context. The context represents all the terrestrial/aerial/satellite user information, indicating spatial-temporal network traffic characteristics for a user-centric network, including geolocation, user behavior and preference, personal trajectory, content popularity, and popular region~\cite{DD-1}. The methods to gain context are named context-aware or context-awareness. The optimization algorithm requires a context-aware module that automatically collects and analyzes data from different sources (e.g., online data and personal devices), then
supplies context for any network management purpose, such as efficiently allocating communication resources while optimizing the aerial nodes' trajectory. Hence, a data-driven ML solution will help to make quick, effective, and improved decisions, especially in
a distributed sense, on network management optimization problems in SEI-VHetNets.
\end{itemize}
\section{Conclusion}
Beyond 5G networks of the future are expected to usher in a radical paradigm shift both in their architecture and network management designs. These highly agile networks can support global coverage and connectivity, and meet the demands of new use cases and applications while efficiently dealing with network complexity. To provide such agility, enabling self-evolving capabilities appears to be inevitable. In this work, we discussed the merits of integrated three-tier vertical architecture, i.e., integrated VHetNet, as a promising solution in support of self-evolving networks. New challenges and difficulties associated with integrated VHetNet architecture were also highlighted from a network management perspective by considering a general scenario.\\ 
\indent Furthermore, in drawing attention to the importance of employing AI/ML methods for efficiently dealing with network management problems and providing full coordination and integration in integrated VHetNets, we discussed the current literature on network management and AI/ML solutions. To ensure that SEI-VHetNets are capable of operating in a fully intelligent, automated, and online way, with full integration and coordination between vertical tiers, we proposed an intelligent framework for  SEI-VHetNets. We also identified the associated core challenges and requirements. Finally, we discussed potential solutions and future research directions to resolve these challenges and fulfill the vision of SEI-VHetNets.

%
%\bibliographystyle{IEEEtran}
%\bibliography{references}

\end{document}